\documentclass[sigplan,10pt]{acmart}

\renewcommand\footnotetextcopyrightpermission[1]{}

\usepackage{tikz}
\usepackage{amsmath}

\usepackage{xcolor}
\usepackage{ipsum}
\usepackage{listings}
\usepackage{caption}
\usepackage{mathrsfs}
\usepackage[T1]{fontenc}

\usepackage{pifont}

\usepackage[linesnumbered,ruled,lined]{algorithm2e}

\usepackage{bm} 

\usepackage[small,compact]{titlesec}

\newcommand{\diffadd}[1]{\colorbox{green!20}{+\ \texttt{\detokenize{#1}}}}
\newcommand{\diffdel}[1]{\colorbox{red!20}{-\ \texttt{\detokenize{#1}}}}

\newcommand{\textop}[1]{\texttt{#1Op}} 

\SetAlgorithmName{Alg.}{}{}

\SetAlFnt{\footnotesize}
\SetAlCapFnt{\sffamily\bfseries\small}
\SetAlCapNameFnt{\unskip\sffamily\bfseries\small}

\lstdefinelanguage{CustomPython}{
  morekeywords={self,cls,__init__,def,return,for,if,else,elif,while,import,from,as,with,pass,continue,break,assert,raise,try,except,finally,global,nonlocal,lambda,yield,del,not,and,or,is,in,not,print,exec,eval,open,close,read,write,append,seek,split,join,strip,replace,find,index,lower,upper,format,join,split,append,extend,remove,pop,insert,sort,reverse,clear,copy,deepcopy},
  sensitive=true,
  morecomment=[l]{\#},
  morestring=[b]",
  morestring=[b]',
}

\lstdefinestyle{diffpython}{
  language=CustomPython,
  basicstyle=\ttfamily\footnotesize\linespread{0.85}\selectfont,
  keywordstyle=\color{teal}\bfseries,
  commentstyle=\color{gray}\slshape,
  stringstyle=\color{purple},
  showstringspaces=false,
  numbers=left,
  numberstyle=\scriptsize\color{gray},
  numbersep=8pt,
  stepnumber=1,
  numberblanklines=false, 
  frame=leftline,
  framesep=4pt,
  xleftmargin=8pt,
  xrightmargin=0pt,
  escapeinside={(*@}{@*)},
  columns=flexible,
  keepspaces=true,
  breaklines=true,
  breakatwhitespace=true,
  breakindent=2pt,
  tabsize=4,
}

\usepackage{graphbox}
\usepackage{fancyvrb}
\usepackage{xspace}
\usepackage{url}
\usepackage{xcomment}
\usepackage[normalem]{ulem}
\usepackage{enumitem}
\usepackage{amsmath}
\usepackage{caption}
\usepackage{subcaption}

\usepackage[font={small,bf,sf}, textfont={small,sf}]{caption}
\usepackage{subcaption}
\usepackage{array}
\usepackage{longtable}
\usepackage{setspace}
\usepackage{listings}
\usepackage{booktabs}
\usepackage{multirow}
\usepackage{pifont}
\usepackage{balance}

\usepackage{mdframed}

\usepackage{soul}

\usepackage{etoolbox}
\makeatletter
\patchcmd{\SOUL@ulunderline}{\dimen@}{\SOUL@dimen}{}{}
\patchcmd{\SOUL@ulunderline}{\dimen@}{\SOUL@dimen}{}{}
\patchcmd{\SOUL@ulunderline}{\dimen@}{\SOUL@dimen}{}{}
\newdimen\SOUL@dimen
\makeatother

\usepackage{footnote}

\usepackage{xcolor}

\soulregister\cite7
\soulregister\ref7
\soulregister\pageref7
\soulregister\eg7
\soulregister\unit7
\soulregister\noteprp7
\soulregister\gap7
\soulregister\mypar7
\soulregister\myparr7
\soulregister\mypari7
\soulregister\myparii7
\soulregister\tinyskip7
\soulregister\code7
\soulregister\footnote7
\soulregister\notemm1
\soulregister\Cref7

\newcommand{\ie}{i.e.,\ }
\newcommand{\eg}{e.g.,\ }

\newcommand{\unit}[1]{\mbox{\hspace{2pt}#1}\xspace}

\newcommand{\Eq}{\mbox{Eq.\hspace{0.25em}}}
\newcommand{\F}{\mbox{Fig.\hspace{0.25em}}}

\newcommand{\A}{\mbox{Alg.\hspace{0.25em}}}

\usepackage{tikz}
\newcommand*\myc[1]{%
\scalebox{0.78}{\begin{tikzpicture}[baseline=-3pt]
  \protect\node[draw,circle,inner sep=0.5pt, fill=black] {\textcolor{white}{\textsf{\textbf{#1}}}};
\end{tikzpicture}}}

\newenvironment{myitemize}{%
\begin{itemize}[leftmargin=1em, itemsep=.1em, parsep=.1em, topsep=.1em,
    partopsep=.1em]}
{\end{itemize}}

\newcommand{\tinyskip}{\vspace{2pt}}

\newcommand{\mypar}[1]{\tinyskip\noindent\textbf{#1.}\xspace}
\newcommand{\myparr}[1]{\tinyskip\noindent\textbf{#1}\xspace}

\makeatletter
\lst@AddToHook{OnEmptyLine}{\addtocounter{lstnumber}{-1}\\[-.3em]}%
\makeatother

\newcommand{\tighttimes}{\mkern-3mu\times\mkern-3mu}
\newcommand{\htighttimes}{\mkern-1mu\times\mkern-1mu}
\newcommand{\sys}{\textsc{Tempo}\xspace}

\begin{document}

\date{}

\title[\sys{}: Compiling Dynamic Deep Learning]{\sys{}: Compiled Dynamic Deep Learning with\\ Symbolic Dependence Graphs}
\author{Pedro F. Silvestre}
\email{p.silvestre21@imperial.ac.uk}
\affiliation{%
  \institution{Imperial College London}
  \country{}
}

\author{Peter Pietzuch}
\email{prp@imperial.ac.uk}
\affiliation{%
  \institution{Imperial College London}
  \country{}
}


\begin{abstract}

Deep learning~(DL) algorithms are often defined in terms of \emph{temporal relationships}: a tensor at one timestep may depend on tensors from earlier or later timesteps. Such \emph{dynamic} dependencies (and corresponding dynamic tensor shapes) are difficult to express and optimize: while \emph{eager} DL systems support such dynamism, they cannot apply compiler-based optimizations; \emph{graph-based} systems require static tensor shapes, which forces users to pad tensors or break-up programs into multiple static graphs.

We describe \sys{}, a new DL system that combines the dynamism of eager execution with the whole-program optimizations of graph-based compilation. \sys{} achieves this through a declarative programming model with \emph{recurrent tensors}, which include explicit \emph{temporal dimensions}. Temporal dimensions can be indexed using \emph{symbolic expressions} to express dynamic dependencies on past and future tensors. Based on this, \sys{} constructs a \emph{symbolic dependence graph}, which concisely encodes dynamic dependencies between operators, and applies whole-program optimizations, such as algebraic simplifications, vectorization, tiling, and fusion. By tiling dynamic dependencies into static-size blocks, \sys{} can also reuse existing static code-generators. It then uses a polyhedral model to find a feasible execution schedule, which includes memory management operations. We show that \sys{} achieves a 7$\times$ speedup over JAX for Llama-3.2-3B decoding; for reinforcement learning algorithms, \sys{} achieves a 54$\times$ speedup, with 16$\times$ lower peak memory usage.

\end{abstract}


\settopmatter{printfolios=true,printacmref=false}
\maketitle
\pagestyle{plain}


\vfill{}
\noindent\textbf{To appear in the Proceedings of the ACM SIGOPS 31st Symposium on Operating Systems Principles (SOSP ’25), October 13–16, 2025, Seoul, Republic of Korea.}

\section{Introduction}

Deep Learning~(DL) has had enormous success in domains such as robotics~\cite{hwangbo2019learning}, games~\cite{silver2017mastering}, natural language~\cite{brown2020language}, protein folding~\cite{jumper2021highly}, and image generation~\cite{rombach2022high}. DL computation is defined using operators~(\eg matrix multiplication, addition, and indexing) with input and output \emph{tensors}. Tensors are multi-dimensional arrays laid out spatially in memory, whose \emph{shape}---a tuple of integers---specifies the size of each spatial dimension. DL workloads are resource-intensive, and DL systems, such as Torch~\cite{paszke2019pytorch} or JAX~\cite{frostig2018compiling}, offer high-level tensor programming models for GPU accelerators~\cite{raina2009large}. 

A common feature of many DL algorithms~\cite{vaswani2017attention, sutton2018reinforcement, rumelhart1986learning, rombach2022high, gu2021efficiently} is their mathematical definition through \emph{recurrence equations}~\cite{karp1967organization}. Recurrence equations express \emph{temporal} relationships between values at different timesteps (past and future) declaratively, \ie without specifying an order of execution. For example, attention~\cite{vaswani2017attention}---a key operation in large language models~(LLMs)---depends, at a given timestep, on outputs from all prior timesteps; reinforcement learning~(RL)~\cite{sutton1988learning, williams1992simple} algorithms produce actions at each timestep and use a sum of future rewards to evaluate them.

Such \emph{dynamic} (\ie time-varying) dependencies between operators and tensors of different timesteps  introduce challenges when existing DL systems try to support them:
\begin{myitemize}
  \item \textbf{Expressiveness}---dynamic dependencies can create tensors with \emph{dynamic shapes} whose support is limited in terms of current APIs~\cite{ansel2024pytorch,lai2025relax}; 
  \item \textbf{Optimization}---existing graph representations~\cite{abadi2016tensorflow, frostig2018compiling} cannot express dynamic dependencies, missing opportunities for algebraic simplification or vectorization;  
  \item \textbf{Scheduling}---future dependencies complicate scheduling, as dependent computations must be delayed~\cite{karp1967organization, moritz2018ray};
  \item \textbf{Memory management}---dependencies on past timesteps result in complex decisions on which tensors to retain, deallocate or evict with limited GPU memory~\cite{rhu2016vdnn}.
\end{myitemize}

\begin{figure*}[t]
  \centering
  \includegraphics[width=1.0\textwidth]{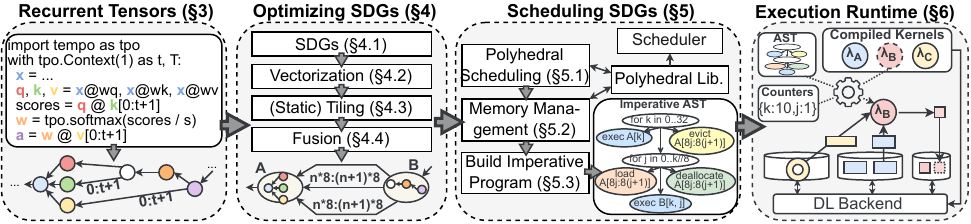}
  \caption{Overview of how \sys{} expresses, optimizes, schedules and manages the memory of dynamic DL algorithms}
  \label{fig:overview}
\end{figure*}

\noindent
\emph{Eager} DL systems, such as Torch~\cite{paszke2019pytorch} or TF-Eager~\cite{agrawal2019tensorflow}, support dynamism through the host language~(Python) by dispatching GPU kernels on-demand~\cite{chetlur2014cudnn}. This imperative approach, however, prevents them from performing compiler-based optimizations~\cite{ansel2024pytorch} or scheduling; in contrast, \emph{graph-based systems}, such as TensorFlow~\cite{abadi2016tensorflow} and JAX~\cite{frostig2018compiling}, enable ahead-of-time compilation by first constructing a \emph{dataflow graph} in which each node represents a DL operator and edges are tensors that flow between them. 
Optimizations such as memory planning~\cite{maas2022telamalloc,liu2024uncovering}, operator fusion~\cite{chen2018tvm}, vectorization~\cite{ragan2013halide}, and code-generation~\cite{Lattner_2021, vasilache2018tensor}, however, require statically known fixed-size tensor shapes.

To compile dynamic DL algorithms, users must \emph{pad} dynamic tensor shapes to a static size, and use \emph{masks} to encode dynamic dependencies, which leads to large memory and execution overheads~\cite{zheng2023bladedisc}. Alternatively, users can break-up computation into static sub-programs, but this prevents whole-program optimizations leading to worse performance~\cite{hessel2021podracer}.

Our key insight is that tensors in dynamic DL algorithms have implicit \emph{temporal dimensions}.
By making them explicit, dynamic dependencies become regular tensor-indexing operations across temporal dimensions. Furthermore, we show that dynamism in dependencies and shapes has structure, and can be concisely represented using \emph{symbolic expressions} of timestep symbols, enabling optimization.

We describe \sys{}, a new DL system that expresses, optimizes, and schedules dynamic DL programs, as well as manages their memory. \sys{} makes the following novel technical contributions (shown in \F\ref{fig:overview}):

\mypar{(1)~Declarative programming with recurrent tensors~(\S\ref{sec:rts})} To capture dynamic dependencies, we introduce \emph{recurrent tensors}~(RTs), a declarative programming model inspired by recurrence equations~\cite{karp1967organization}, often used in DL algorithms~\cite{schulman2015high, vaswani2017attention, sutton1988learning, rumelhart1986learning}. RTs can have multiple temporal dimensions (\eg timestep and iteration), which define a temporal \emph{domain}. RTs can be indexed using symbolic expressions to declaratively express dynamic dependencies. RTs have well-defined domain broadcast~\cite{van2011numpy} and backpropagation~\cite{baydin2018automatic} semantics. 

\mypar{(2)~Optimization using symbolic dependence graphs~(\S\ref{sec:pdg_and_transformations})} \sys translates RTs into a \emph{symbolic dependence graph~(SDG)} in which each vertex operator carries the temporal domain of the RT that created it, \ie the set of timesteps at which the operator produces output tensors; edges describe how operators at different timesteps \emph{depend on} each other's outputs using the symbolic expressions extracted from RTs. 

SDGs enable \sys{} to extend classic compiler optimizations (\eg algebraic simplifications, redundant code elimination) to dynamic computations by manipulating symbolic expressions. In addition, SDGs simplify advanced transformations, such as \emph{vectorization} (batching many small operations) and \emph{tiling} (executing large operations in smaller tiles), by viewing them as moving tensor dimensions from temporal to spatial and vice versa. For efficient code-generation, \sys{} tiles dynamic dependencies into static-size tiles and \emph{fuses} islands of static operations into single operations.

\mypar{(3)~Scheduling using a polyhedral model~(\S\ref{sec:polyhedral_model})} The declarative RT program does not prescribe an execution order, and, due to their dynamic dependencies, SDGs cannot be scheduled using  topological sorting. Instead, \sys{} uses the polyhedral model~\cite{verdoolaege2010isl, bondhugula2008pluto} for program-wide scheduling: \sys{} translates the SDG into a set of constraints and solves an integer linear program to obtain a schedule that assigns a physical execution time to each timestep. Based on the schedule, \sys{} generates an imperative program for the user computation, represented as an abstract syntax tree~(AST).

\mypar{(4)~Automated memory management~(\S\ref{sec:mem_sched})} \sys{} introduces a simple method for automatic memory management, including buffer deallocations, donations, and GPU/CPU buffer swapping. It augments the SDG with explicit memory management operations, which are restricted by dependency edges. Therefore, memory management is also handled when solving for an execution schedule.

\tinyskip

\noindent
Our prototype implementation of \sys{}\footnote{Available  at \url{https://github.com/LSDS/Tempo}} supports single-GPU execution on multiple execution engines (Torch and JAX). To execute SDGs, \sys{} picks physical tensor memory representations based on dependence expressions and applies existing code-generators~\cite{openxla2024} to fused operators.
It then follows the schedule AST, evaluating symbolic dependence expressions, to resolve dynamic dependencies at runtime.

Our evaluation shows that \sys{} effectively compiles dynamic DL programs, while automating algorithm-specific scheduling and memory management. For LLM decoding using Llama-3.2-3B~\cite{dubey2024llama}, we compare \sys{} against Torch and JAX and show that it is up to 7$\times$ faster and scales to 4$\times$ longer sequences due to its awareness of dynamic dependencies. For three typical RL algorithms~\cite{schulman2017proximal, williams1992simple, sutton1988learning} with different dynamic dependencies, \sys achieves up to 54$\times$ faster end-to-end training times than five baseline RL systems by eliminating redundant computation. For both workloads, \sys{} has an up to 16$\times$ lower peak GPU memory usage due to its tiling optimization and efficient memory management.



\section{Supporting Dynamic DL Programs}\label{sec:bg}

  \begin{figure}[t]
  \begin{subfigure}[t]{.475\columnwidth}
    \centering
    \includegraphics[width=\columnwidth]{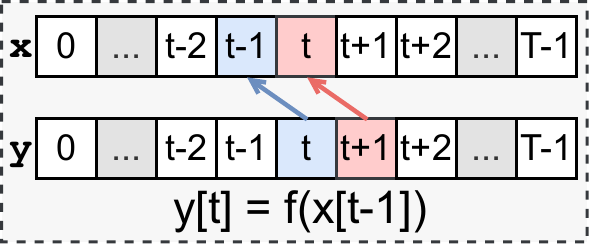}
    \caption{State-passing~\textnormal{\cite{rumelhart1986learning,gu2021efficiently,ho2020denoising}}}\label{fig:access_patterns:state_passing}
  \end{subfigure}
  \hfill
  \begin{subfigure}[t]{.475\columnwidth}
    \centering
    \includegraphics[width=\columnwidth]{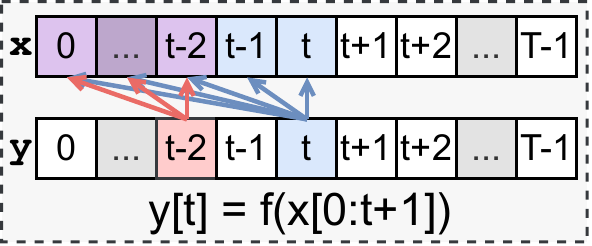}
    \caption{Causal~\textnormal{\cite{vaswani2017attention, ouyang2022training}}}\label{fig:access_patterns:causal}
  \end{subfigure}
  
  \begin{subfigure}[t]{.475\columnwidth}
    \centering
    \includegraphics[width=\columnwidth]{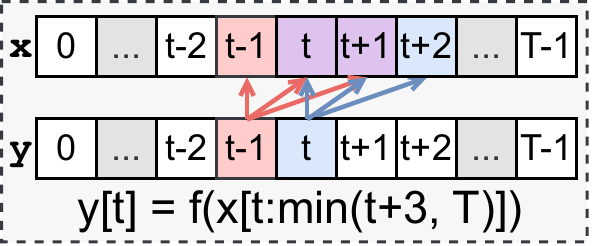}
    \caption{Forward window~\textnormal{\cite{mnih2016asynchronous, sutton1988learning}}}\label{fig:access_patterns:fwd_window}
  \end{subfigure}%
  \hfill
  \begin{subfigure}[t]{.475\columnwidth}
    \centering
    \includegraphics[width=\columnwidth]{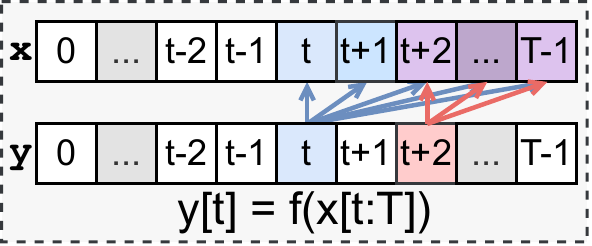}
    \caption{Anti-causal~\textnormal{\cite{williams1992simple, ouyang2022training, schulman2017proximal}}}\label{fig:access_patterns:anti_causal}
  \end{subfigure}%

  \begin{subfigure}[t]{.475\columnwidth}
    \centering
    \includegraphics[width=\columnwidth]{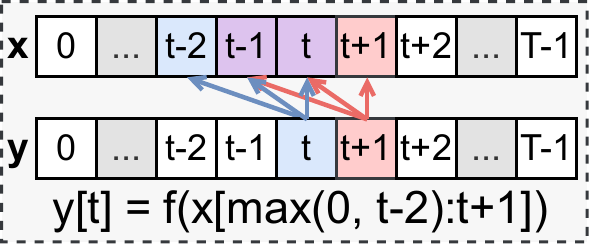}
    \caption{Backward window~\textnormal{\cite{zaheer2020big, mnih2013playing}}}\label{fig:access_patterns:bwd_window}
  \end{subfigure}
  \hfill
  \begin{subfigure}[t]{.475\columnwidth}
    \centering
    \includegraphics[width=\columnwidth]{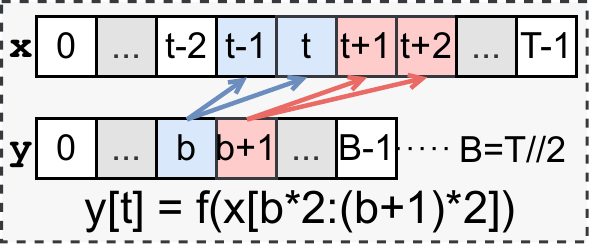}
    \caption{Block-based~\textnormal{\cite{schulman2017proximal, lea2017temporal, ma2022mega}}}\label{fig:access_patterns:block}
  \end{subfigure}%

  \caption{Dynamic dependencies \textnormal{(Tensor~$y$ at time~$t$ depends on (as indicated by the arrow direction) a dynamic range of $x$~values.)} }
  \label{fig:access_patterns}
\end{figure}

Today's DL algorithms include diverse dynamic dependencies, often described in papers using recurrence equations~\cite{karp1967organization}, such as those in \F\ref{fig:access_patterns}.
For example, recurrent~\cite{rumelhart1986learning}, state-space~\cite{gu2021efficiently}, and diffusion models~\cite{ho2020denoising} all pass state from a timestep to the next~(\F\ref{fig:access_patterns:state_passing}, where $x$ may equal $y$).
DQN~\cite{mnih2013playing} and windowed attention~\cite{zaheer2020big} use backwards windows~(\F\ref{fig:access_patterns:bwd_window}).
PPO~\cite{schulman2017proximal}, video models~\cite{ma2022mega}, and long-context transformers~\cite{dai2019transformer}  access data in chunks~(\F\ref{fig:access_patterns:block}).
We now focus on two popular workloads and their dynamic dependencies.

\begin{figure}[t]
  \centering
  \includegraphics[width=\columnwidth, trim={1 0 2 4}, clip]{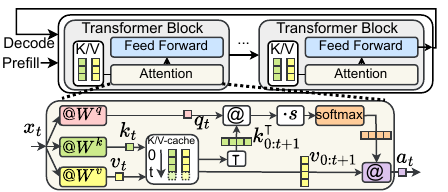}
  \caption{Simplified Llama-3.2-3B architecture  \textnormal{(Each decoding step depends on a dynamic range of cached key-value (K/V) pairs.)}}
  \label{fig:transformer_decoding}
\end{figure}

\myparr{Large language models~(LLMs)}~\cite{brown2020language, dubey2024llama} generate text as sequences of tokens. \F\ref{fig:transformer_decoding} shows a transformer-based LLM architecture. After an initial \emph{user prompt} processing step (prefill), the LLM begins iteratively \emph{decoding}, \ie generating a response one token at a time, until it generates a special \emph{end-of-sequence} token at time $T$.  LLMs are stacks of transformer blocks, each with a feed-forward and an \emph{attention} layer.

Attention uses dynamic computation. It first projects the input~$x_t$ into query $q_t$, key $k_t$, and value $v_t$ vectors, and 
then computes attention (\Eq\ref{eq:attention}),
where $s$ is a scaling constant and $\phi(t)$ is called the \emph{attention pattern}.
This pattern defines a dynamic range of keys and values used to compute attention at timestep~$t$.
Typical patterns include causal attention~\cite{vaswani2017attention}~(\F\ref{fig:access_patterns:causal}) where $\phi(t)=0\!:\!t\!+\!1$,
and window attention~\cite{zaheer2020big}~(\F\ref{fig:access_patterns:bwd_window}) where $\phi(t)=\textrm{max}(t\!-\!w, 0)\!:\!t\!+\!1$, with window size~$w$.
The dynamic section ends when the dot product with values contracts the dynamically-sized dimension.
\begin{eqnarray}
    a_t = \textrm{softmax}(q_t k_{\phi(t)}^{\intercal} \cdot s) \, v_{\phi(t)} \label{eq:attention}
\end{eqnarray}

\begin{figure}[t]
  \centering
  \includegraphics[width=0.975\columnwidth, trim={12 2 9 0}, clip]{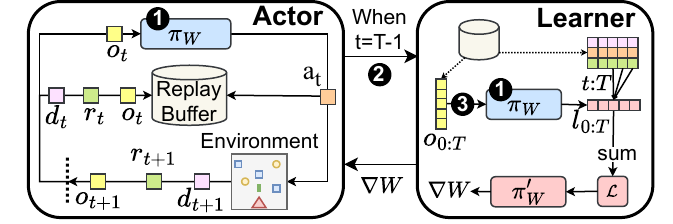}
  \caption{Actor-learner architecture in RL systems \textnormal{(Decoupling acting and learning causes \ding{202} duplicate forward passes, \ding{203} serial acting and learning, and \ding{204} high peak memory usage.)}}
  \label{fig:actor_learner}
\end{figure}

\myparr{Reinforcement learning~(RL)}~\cite{sutton2018reinforcement} learns an \emph{acting policy}~$\pi_{W}$, represented by a deep neural network~(DNN) with parameters~$W$, from \emph{experience} data collected by trial-and-error interactions with a simulated \emph{environment} (see~\F\ref{fig:actor_learner}).
At each \emph{timestep}~$t$, starting from the initial observation~$o_0$, the \emph{acting policy} produces an action~$a_t$, which is executed in the environment and, in response, receives the next observation~$o_{t+1}$, a done signal~$d_{t}$, and a reward~$r_{t}$.

In each iteration, a \emph{learner} computes a loss~$\mathcal{L}$ from the generated experience data and backpropagates it into a parameter update~$\nabla_W$. 
The loss is a sum of per-timestep losses, $\mathcal{L} = \sum_{t=0}^T l_{t}$,
where each $l_{t}$ \emph{dynamically} accesses only a \emph{portion} of the experience data, depending on the algorithm. 
REINFORCE~\cite{williams1992simple} accesses all future steps~(\F\ref{fig:access_patterns:anti_causal}), while A2C~\cite{mnih2016asynchronous} uses a forward-looking window~(\F\ref{fig:access_patterns:fwd_window}).

\subsection{Dynamic DL and whole-program compilation}
\label{sec:dataflow}

Dynamic algorithms are difficult to represent and optimize in today's DL systems, preventing whole-program compilation.
\emph{Imperative (eager)} systems, such as PyTorch~\cite{paszke2019pytorch} and TF-Eager~\cite{agrawal2019tensorflow}, support dynamism well: users write imperative code that dynamically dispatches GPU kernels that are appropriately-sized to the dynamic shape~\cite{chetlur2014cudnn}. 
They rely on the host language (\eg Python) to evaluate control-flow, conditionals, and dynamic dependencies. 
These systems, however, lack a graph representation that can be optimized and scheduled ahead-of-time, thus placing the burden of writing efficient code on the user. \emph{Lazy eager} approaches~\cite{suhan2021lazytensor} collect small subgraphs before optimizing them, but this is still insufficient for whole-program compilation. 

\emph{Declarative (graph-based)} systems, such as TensorFlow~\cite{abadi2016tensorflow} and JAX~\cite{frostig2018compiling}, first build a dataflow graph of the program before performing optimizations and execution scheduling. TensorFlow uses a flat dataflow graph with several control-flow operators~\cite{yu2018dynamic} that encode loops and conditionals; JAX uses a single \texttt{while} operator~\cite{xla_op_semantics}, with nested subgraphs for loop body and condition. Both approaches, however, require static tensor shapes in the graph, and thus cannot directly express dynamic data dependencies.

As a result, users are forced into different workarounds to implement dynamic computation, 
each with drawbacks:

\mypar{(a)~Per-shape compilation} Users can compile a separate graph for each possible tensor shape, but this is expensive for highly variable shapes~\cite{suhan2021lazytensor} as found in LLMs and RL.

\mypar{(b)~Unrolling loops} Users can statically unroll the timestep loop~\cite{frostig2018compiling} when $T$ is known, generating all instantiations of dynamic dependencies. This leads to \emph{graph size explosion}, increasing compilation times, and also loses the semantics of dynamic dependencies, which prevents optimizations.

\mypar{(c)~Padding and masking} Users can coarsen dynamic dependencies, padding them to a fixed maximum sequence length $T'>T$. For correctness, users must also mask out valid timesteps. This is cumbersome and introduces processing and memory overheads.

\subsection{Consequences in DL implementations}
\label{sec:frameworks}

We now discuss how the lack of support for dynamic dependencies affects the implementations of DL algorithms.

\tinyskip

\noindent

In \textbf{LLM inference}, the dependence on past key/value~(K/V) pairs hinders whole-program compilation. This dependence creates dynamic tensor shapes, which require inefficient padding to compile. Furthermore, to avoid recomputing keys and values, K/V caches are used~(see~\F\ref{fig:transformer_decoding}), often managed by external systems~\cite{kwon2023efficient,qin2024mooncake}. 

Managing these caches is complex because different types of attention require different cache policies. 
While causal attention~\cite{vaswani2017attention} must store all prior K/V pairs, window attention~\cite{zaheer2020big} allows for the deallocation of old K/V pairs. Linear~\cite{katharopoulos2020transformers}, global~\cite{zaheer2020big}, block~\cite{dai2019transformer}, and full attention~\cite{devlin2019bert}, or hybrid models that interleave attention types~\cite{team2024gemma}, demand even more nuanced cache policies.
Since existing LLM inference systems~\cite{kwon2023efficient, zheng2024sglang} do not model dynamic dependencies, they struggle to offer tailored cache policies,\footnote{\url{https://github.com/vllm-project/vllm/issues/15705}} requiring large engineering effort for each new supported attention type.\footnote{\url{https://github.com/vllm-project/vllm/pull/14097}}

\myparr{RL frameworks}~\cite{stooke2019rlpyt, Petrenko2020SampleFE, hoffman2020acme, liang2018rllib, weng2021tianshou, rl-games2022, bou2023torchrl} address forward-looking dependencies in loss computation by coarsening them to $0\!:\!T$ and decomposing the computation into two subgraphs: (i)~an \emph{actor} graph executed repeatedly until termination, which produces the full simulation data and stores results in a \emph{replay buffer}~\cite{cassirer2021reverb, wang2023gear}; and (ii)~a \emph{learner} graph that is invoked once after the simulation terminates to compute a parameter update using the experience from the replay buffer.

While the actor-learner model can be applied to most RL algorithms, it has several drawbacks (indicated by the numbers in \F\ref{fig:actor_learner}) due to the independently compiled subgraphs:

\mypar{(1)~Redundant computation} Intermediate activations produced by the actor~$\pi_{W}$ are discarded and recomputed in the learner for use in back-propagation. This adds substantial overhead, as it doubles the amount of inference work per timestep for typical algorithms; 

\mypar{(2)~Missed parallelism} The execution of actors and learners must be serialized due to the coarse $0\!:\!T$ dependency. This prevents the parallelization of learning on past timesteps while acting out future timesteps; and

\mypar{(3)~High peak memory usage} The ``all-at-once'' learning prevents early deallocations and results in high peak GPU memory usage, potentially causing out-of-memory errors.\footnote{Consider learning from small $W\!=\!H\!=\!256$ images with $C\!=\!3$ color channels, at $P\!=\!4$b float precision, with a batch size of $B\!=\!512$, and $T\!=\!1000$ timesteps. This requires  at least $T\!\times\!B\!\times\!(W\!\times\!H\!\times\!C\!\times\!P) / 2^{30}=$ 375\unit{GB} of memory.}




\section{Recurrent Tensors}
\label{sec:rts}

\begin{figure}[t]
  \centering
  \includegraphics[width=\columnwidth, trim={1.6 1 1 2}, clip]{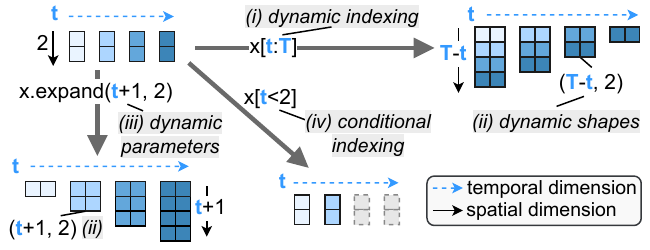}
  \caption{How a RT \boldmath{$x$} with domain \boldmath{$(t,)$} can be used in \sys{}}
  \label{fig:rts}
\end{figure}

While writing dynamic algorithms is challenging in current DL systems (see~\S\ref{sec:dataflow}), we observe that they are naturally expressible as \emph{recurrence equations}~\cite{karp1967organization}. This makes recurrence equations ubiquitous in algorithmic pseudocode, \eg when defining optimizers~\cite{kingma2014adam}, learning rate schedules~\cite{darken1990note}, models~\cite{werbos1990backpropagation, bengio2021flow, ho2020denoising}, layers~\cite{vaswani2017attention,zaheer2020big}, and loss functions~\cite{schulman2015high, schulman2017proximal}.

%
%

To exploit this expressiveness, \sys{} introduces \emph{recurrent tensors~(RTs)}. In addition to the spatial dimensions for its shape, an RT has \emph{temporal dimensions}, such as iterations or timesteps, which define its \emph{domain}~(see~\F\ref{fig:rts}). 
Each temporal dimension has an associated pair of symbols---a \emph{current step}~(\eg~$t$,~$i$) and an \emph{upper bound}~(\eg~$T$,~$I$).
If an RT $x$ has $t$ in its domain, it is defined by a recurrence equation for every point $0 \leq t < T$, and its value (and shape) may vary with $t$.
RTs enable declarative programming, as their defining equations can refer to past or future values of other RTs by indexing them with \emph{symbolic expressions}.


\myparr{Symbolic expressions} are constructed from symbols using overloaded operators, such as arithmetic (\eg $+$, $/$), boolean comparison (\eg $\leq$, $=$) and logic (\eg $\&$, $|$), as well as $\text{min}$ and $\text{max}$. Accessing a range of temporal values is done using a \emph{slice expression} ({$\mathrm{start}\!:\!\mathrm{stop}\!:\!\mathrm{step}$}), and to express multiple dimensions at once, \emph{sequence expressions} $([\cdot,\cdot])$ are used.

Symbolic expressions are versatile and can be used to~(as shown by the labels in \F\ref{fig:rts}):~(i)~index the temporal dimensions of RTs expressing dynamic dependencies;~(ii)~express dynamic shapes;~(iii)~parameterize tensor operations which receive index or shape arguments (\eg a dynamic expand or spatial index select);~and~(iv)~naturally express conditional computation using boolean expressions. 
Indexing an RT’s temporal dimension with a slice expression creates a (possibly dynamic) leading spatial dimension in the output RT.

\begin{algorithm}[t]
  \caption{REINFORCE using RTs. \textnormal{(Line 13 is discussed later.)}}
  \label{alg:experimentation}
\begin{lstlisting}[style=diffpython, linewidth=\columnwidth]
  import tempo as tpo
  ctx = tpo.Context(num_dims=3)
  with ctx as (b, B), (i, I), (t, T):
    env = tpo.rl.env.make("gym.CartPole-v1")
    dnn = tpo.DNNBuilder(domain=(i,))
      .from_env(env, hidden=[32, 32]).build()

    o = tpo.like(env.obs_space, domain=(b, i, t))
    o[b, i, 0] = env.reset(domain=(b, i))
    a = dnn(obs) # Acting
    o[b, i, t+1], r, d = env.step(a)

    # Different access pattern -> different schedule
(*@\diffdel{     g = r[b, i, t:T].discounted_sum(0.95)          }@*)
(*@\diffadd{     g = r[b, i, t:min(t+5, T)].discounted_sum(0.95)}@*)
    l = -dnn.log_prob(a) * g # Learning 
    l[0:B, i, 0:T].mean().backward() 
    lr = 1e-3 * (0.99 ** i) # Decaying learning rate
    tpo.optim.Adam(dnn.params, lr=lr).step()

    dnn[(i+1) % 5 == 0].checkpoint(save_path)
    ctx.run({B: 64, I: 20, T: 200})
\end{lstlisting}
\end{algorithm}

\mypar{REINFORCE with RTs} \A\ref{alg:experimentation} shows an example of the REINFORCE RL algorithm~\cite{williams1992simple} written in \sys. Users begin by setting up a context~(line~2) with a chosen number of temporal dimensions\footnote{With zero temporal dimensions, \sys behaves like a typical DL system with only spatial tensor dimensions.}, associating each dimension with a semantic meaning~(line~3): batch sample~$b$, iteration~$i$, and timestep~$t$. An RL environment is created~(line~4), and a DNN is defined~(line~5). Similar to PyTorch~\cite{paszke2019pytorch}, DNNs are defined by composing modules, but the domain over which parameters vary (iterations) must be explicit~(line~5). 

The observations tensor is defined as a \emph{branching} RT: the first timestep is given by the environment reset~(line~8) and subsequent timesteps are computed by stepping the environment with the DNN's action~(line~10). 
The \emph{returns}~$g$ of each timestep~$t$ are computed from the rewards~$r$ indexed by~\mbox{$[t\!:\!T]$}~(line~12), yielding an RT with \emph{dynamic shape}~\mbox{$(T\!-\!t,)$}. \sys{} propagates dynamic shapes transparently.

The loss is then backpropagated~(lines~14--15), and the gradients are used by the Adam optimizer~\cite{kingma2014adam} (line~17), also defined using RTs, to update the DNN. 
Expressing conditional computation, such as checkpointing~(line~18), is done by indexing with boolean expressions, which only execute when the boolean evaluates to true.
Finally, the program is compiled and run, with the user providing bounds for each temporal dimension (line~19). 

Note how, with RTs, there is no need to decouple acting and learning. The original forward pass can simply be used within the loss definition. Furthermore, there is no need to use replay buffers~\cite{cassirer2021reverb} to store prior timesteps of experience\footnote{A replay buffer is still needed for asynchronous~\cite{mnih2016asynchronous} or offline RL~\cite{levine2020offline}.}, because \sys{} manages the storage of each RT based on the dynamic dependencies. 

\begin{figure}[t]
  \centering
  \includegraphics[width=0.975\columnwidth, trim={0 34 0 0}, clip]{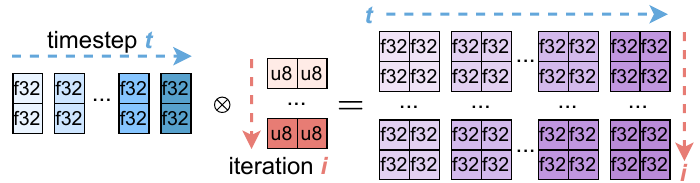}
  \caption{Domain unioning \textnormal{(Similar to shape broadcasting and datatype promotion, domains are unioned when tensors interact.)}}
  \label{fig:domain_unioning}
\end{figure}

\mypar{Automatic domain inference} Annotating domains is often unnecessary, as \sys{} infers them. For example, temporally indexing an RT with a constant (\eg $5$ or $0\!:\!T$) on dimension~$t$ removes $t$ from the resulting RT's domain.
Spatially indexing an RT with $t$ adds $t$ to the resulting RT's domain.
If an RT is not explicitly temporally indexed ($g$ in line~14), it is treated as if indexed by the identity~(\ie~$g[b, i, t]$).

When two RTs interact, their domains are \emph{unioned}, as shown in \F\ref{fig:domain_unioning}. 
In this example, 
the left RT varies with timesteps $t$, while the right RT varies with iterations $i$. Their interaction produces an RT that varies with both (\ie has a domain of $(i,t)$). 
Domain inference allows, \eg for learning rate schedules, to be easily expressed as RTs (line~16), and be integrated with the optimizer without code changes. 

\begin{figure}[t]
  \centering
  \includegraphics[width=0.95\columnwidth, trim={16 1 1 0}, clip]{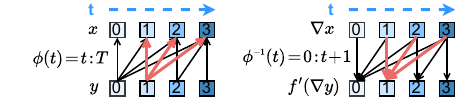}
  \caption{Symbolic automatic differentiation \textnormal{($y[1]$ depends on $x[1\!:\!4]$ and thus $\nabla y[1]$ contributes to $\nabla x[1\!:\!4]$.)}}
  \label{fig:autodiff}
\end{figure}
\mypar{Symbolic automatic differentiation} When \texttt{backward()} is called on an RT, automatic differentiation~\cite{baydin2018automatic} is used to compute the input gradients for an operator based on its output gradients. \sys, however, must also accumulate gradients through temporal dimensions. Consider the RT $y[t] = f(x[\phi(t)])$, where $\phi$ is called the \emph{dependence expression} $y$ has on $x$. A single timestep of $x$ may contribute to several timesteps of $y$.
Thus, the gradient of~$x$, $\nabla x$, is an RT with the same shape, datatype, and temporal domain as $x$, which sums the gradient contributions from all timesteps of $\nabla y$ to which $x[t]$ contributed to:
$$\nabla x[t] = \sum\nolimits_{t' \in \phi^{-1}(t)} f'(\nabla y[t'])$$
\noindent
where $f'$ is the derivative of~$f$, and $\phi^{-1}$ is the \emph{inverse} of the dependence expression. 
Inverting a dependence can be thought of as converting ``which source timesteps does the sink depend on at timestep $t$?'' to ``which sink timesteps depend on the source's timestep~$t$?''. For example, $[t\!+\!3]$ inverts into $[t\!-\!3]$, $[\mathrm{max}(0,t\!-\!3)\!:\!t\!+\!1]$ into $[t\!:\!\mathrm{min}(t\!+\!4, T)]$, and $[t\!:\!T]$ into $[0\!:\!t\!+\!1]$. This last example is depicted in \F\ref{fig:autodiff}.
If $y$ indexes $x$ with a constant expression~(\eg $y = f(x[0\!:\!T])$, $y$ does not vary with $t$. To recover $\nabla x$'s domain, \sys{} must select the corresponding spatial dimension of $y$ using $t$ as an index, \ie $\nabla x[t]\!=\!f'(\nabla y[]).\mathrm{index}(t)$. This process generalizes to multiple temporal dimensions by concatenation.



\section{Symbolic Dependence Graphs}
\label{sec:pdg_and_transformations}

As users define DL programs with RTs, \sys constructs a \emph{symbolic dependence graph~(SDG)} of the program, which encodes the dynamic dependencies between operators.

\subsection{SDG representation}
\label{sec:pdg}

An SDG is a directed graph of \emph{operators}. Each operator may have multiple input or output tensors, and cycles are allowed. 
Each operator~$o$ is tagged with a temporal domain~$\Omega(o)$: the set of temporal \emph{points} in $\mathbb{Z}^{n}$ at which the operator executes, producing output tensors. 

Edges describe how operators at different temporal points depend on each other’s outputs.
Each \emph{edge}~$o_1\xrightarrow{\smash{\phi}}o_2$ is annotated with a \emph{dependence expression}~$\phi$. It indicates that $o_1$ (the sink) at point $p \in \Omega(o_1)$ \emph{depends on} $o_2$ (the source) at points $\phi(p)$. 
The domain~$\Omega$ and dependence expressions~$\phi$ are obtained directly from the RTs that defined them.\footnote{Throughout the paper, to simplify notation, we use $t$ for generic temporal dimensions and  omit \emph{identity} dependencies (\ie $\phi(t)=t$) in figures.}

\mypar{Tensor operators} \sys{} supports a minimal set of 44~stateless operators, most of which are typical elementwise maps, reductions, scans, view, or spatial indexing operators. View and indexing operators (\eg \textop{Reshape}, \textop{Slice}) may use symbolic expressions in shape and index parameters. 
\textop{UDF}s allow users to register custom operations, useful for injecting (retrieving) values to (from) the runtime. \textop{UDF}s may access \emph{external state}. \sys{} uses \textop{UDF}s to integrate with RL environments~\cite{brockman2016openai} and LLM tokenizers~\cite{kudo2018sentencepiece}.

To support branching RTs, we define \textop{Merge}s. They conditionally select which of their multiple inputs to \emph{move} to their single output based on which edge condition~$\psi$ first evaluates to $\mathrm{true}$ at time~$t$. Edge conditions are obtained from the conditional indexing of RTs.

\begin{figure}[t]
  \centering
  \includegraphics[width=1\columnwidth, trim={0 1 1 1}, clip]{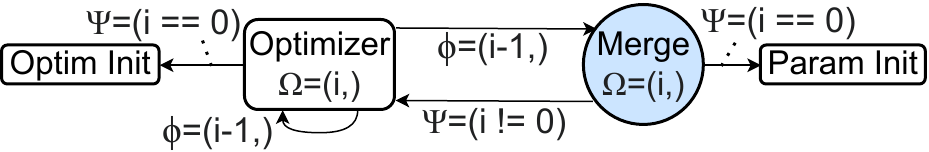}
  \caption{Encoding DNN state using cycles \textnormal{(Boxes are subgraphs.)}}
  \label{fig:params_state}
\end{figure}

\mypar{Encoding state}
To encode state (\eg DNN parameters) in a graph, Tensorflow~\cite{abadi2016tensorflow} uses stateful operators (\texttt{tf.Variable}). JAX~\cite{frostig2018compiling} graphs are stateless, and JAX offloads state management responsibilities to the user.
In \sys{}, state is instead encoded in the SDG using \textop{Merge} cycles.

\F\ref{fig:params_state} shows an example of this for a DNN parameter. The parameter's initial value is given by a parameter initialization subgraph. Subsequent parameter values are obtained from the optimizer step subgraph, which itself depends on the previous parameter value to compute an update.
Optimizers such as Adam~\cite{kingma2014adam} also hold state, which is encoded using the same mechanism.
This approach enables \sys{} to represent state implicitly, without additional operator types that have complex semantics and side-effects, and without requiring users to manage state explicitly.

\mypar{Optimizing symbolically}
\sys{} applies compiler optimizations to SDGs, such as dead and duplicate code elimination, algebraic equivalences, and broadcasting removal,
which are extended to support dynamic dependencies.

Suppose that \sys considers a re-association optimization for a chain of matrix multiplications $(A \cdot B) \cdot C$, where $A\in\mathbb{R}^{32, 16}$, $B\in\mathbb{R}^{16, t}$, and $C\in\mathbb{R}^{t, 8}$. Since $t$ is in the shape of $B$ and $C$, $(A \cdot B) \cdot C$ has a dynamic cost of $32 \cdot 16 \cdot t + 32 \cdot t \cdot 8 = 768 \cdot t$ FLOPs. Re-associating as $A \cdot (B \cdot C)$ costs $16 \cdot t \cdot 8 + 32 \cdot 16 \cdot 8 = 128 \cdot t + 4096$, which is lower when $t > 6.4$.
\sys{} can use the original association for~$t \leq 6$, and re-associate when~$t > 6$.

\sys{} also performs \emph{domain reduction}, which reduces the domain of operators to the union of the domains of their dependencies, thus removing redundant work. This is akin to \emph{loop-invariant code motion}~\cite{kennedy2001optimizing} in optimizing compilers.

\begin{figure}[t]
  \centering
  \begin{subfigure}[t]{0.3\columnwidth}
    \centering
    \includegraphics[width=0.97\columnwidth, trim={1 1 1 1}, clip]{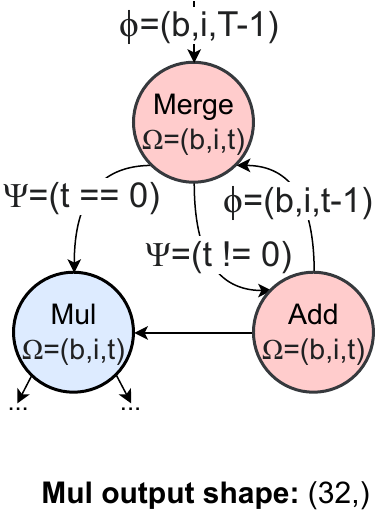}
    \caption{Initial SDG}\label{fig:transformation_pipeline:pdg}
  \end{subfigure}%
  ~
  \begin{subfigure}[t]{0.65\columnwidth}
    \centering
    \includegraphics[width=0.99\columnwidth, trim={1 1 1 1}, clip]{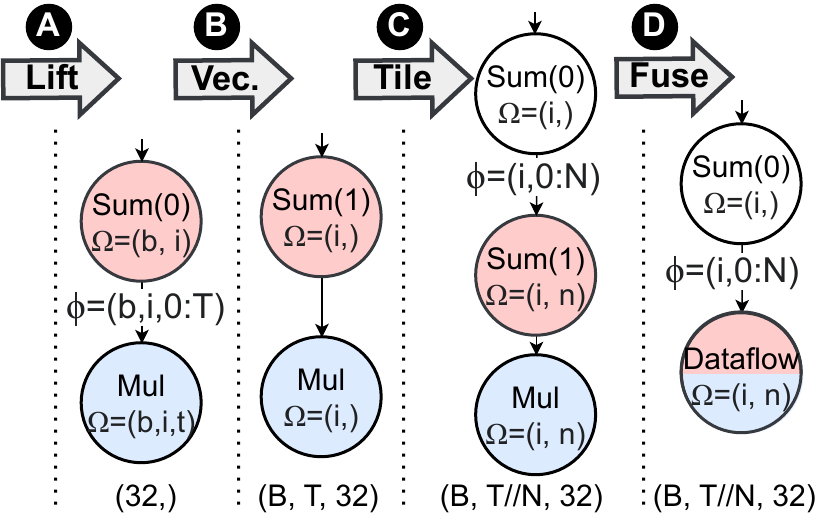}
    \caption{Transformation pipeline}\label{fig:transformation_pipeline:transforms}
  \end{subfigure}%
  \caption{SDG passing through \sys{}'s transformation pipeline}
  \label{fig:transformation_pipeline}
\end{figure}

\begin{figure}[t]
  \centering
  \includegraphics[width=0.95\columnwidth, trim={1 0 2 2}, clip]{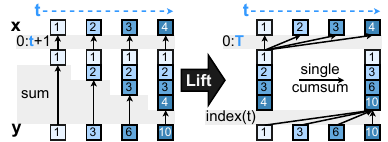}
  \caption{Lifting shown with RTs \textnormal{(left: redundant sum of $0\!:\!t\!+\!1$ at each $t$; right: equivalent with single cumulative sum of $0\!:\!T$)}}
  \label{fig:lift_visual}
\end{figure}

The SDG in \F\ref{fig:transformation_pipeline:pdg} computes the sum of the \textop{Mul}'s output over timesteps $t$ using a \textop{Merge}.
 At $t\!=\!0$, the \textop{Merge} simply copies the \textop{Mul}’s output; otherwise, it adds the current \textop{Mul} output to the sum at $t\!-\!1$. Though correct, this SDG is inefficient. 
\F\ref{fig:transformation_pipeline:transforms} shows how \sys{} progressively \emph{transforms} this SDG for efficient execution.

\mypar{Lifting}
\sys{} starts by
eliminating \emph{recurrent patterns} (reductions, scans~\cite{blelloch1990prefix}, and stencils~\cite{ragan2013halide} implemented using \textop{Merge}s) that prevent vectorization. SDGs enable simple pattern-matching of these complex structures, allowing \sys{} to replace them with batch
\textop{Sum}s, \textop{CumSum}s,  or \textop{Conv}s.
In \F\ref{fig:transformation_pipeline:transforms}, \sys{} \emph{lifts}~\myc{A} the recurrent addition into a \textop{Sum}.

Lifting can also optimize computation. In \F\ref{fig:lift_visual}, the original computation, $y[t]\!=\!x[0\!:\!t\!+\!1]\mathrm{.sum}()$, redundantly sums the prior steps of $x$ at each timestep~$t$. \sys{} replaces this with $y[t]\!=\!x[0\!:\!T]\mathrm{.cumsum()}$, and restores the temporal dimension $t$ using a spatial indexing operation with $t$ as an index.

\subsection{Vectorization}
\label{sec:vectorization}

\emph{Vectorization}~(\myc{B} in \F\ref{fig:transformation_pipeline:transforms}) reduces the execution time of SDGs by laying out temporal dimensions spatially. Vectorized operators execute fewer times but with larger input and output tensors. This offers greater SIMD parallelism~\cite{flynn2005very}, at the cost of increased peak memory usage.

For each temporal dimension $t$, \sys{} finds a set of vectorizable operators and applies a \emph{vectorization rule} to each. 
Such rules are simple: most operators simply have $t$ removed from their domain and $T$ prepended to their input and output shapes; operators with dimension parameters (\eg \textop{Sum}, \textop{Squeeze}) have them incremented, to account for $T$ being prepended to their input shape; and operators with shape parameters (\eg \textop{Reshape}, \textop{Expand}) also prepend $T$ to this shape.
When vectorizing \textop{Index}s or \textop{IndexAdd}s and both their source (\ie the tensor being indexed) and index input tensors are also being vectorized, \sys{} must promote these operators into \textop{Gather} and \textop{ScatterAdd}, because index operations only support one-dimensional indexes.

\mypar{Vectorizability} 
Cycles complicate vectorization, because incorrectly vectorizing them can lead to \emph{unschedulable} SDGs. 
\sys{} avoids unschedulable SDGs using the following conservative approach: we define a cycle as \emph{trivial} on $t$ if all symbolic indexes of that dimension are $[t]$ itself or a constant expression (\eg $[0:T]$, $[5]$, $[0:5]$). \sys does not vectorize operators in non-trivial cycles; operators in trivial cycles must all be vectorized or none can be.

\begin{figure}[t]
  \begin{subfigure}[t]{.28\columnwidth}
    \centering
    \includegraphics[width=\textwidth]{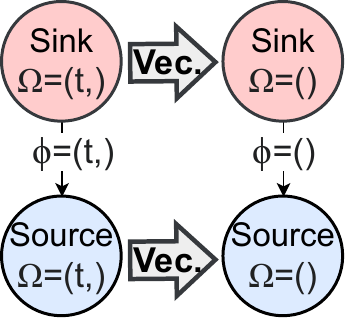}
    \caption{Both}\label{fig:vec:vec_both}
  \end{subfigure}
  \hfill
  \begin{subfigure}[t]{.42\columnwidth}
    \centering
    \includegraphics[width=0.665\textwidth]{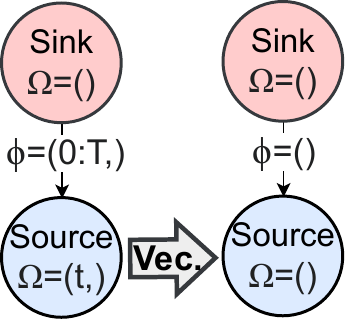}
    \caption{Source~$(t \notin \Omega(\textnormal{sink}))$}\label{fig:vec:vec_src_sink_no_dim}
  \end{subfigure}
  \hfill
  \begin{subfigure}[t]{.28\columnwidth}
    \centering
    \includegraphics[width=\textwidth]{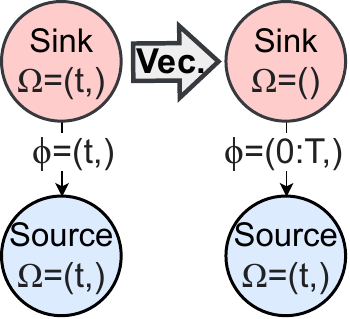}
    \caption{Sink}\label{fig:vec:vec_sink_src_has_dim}
  \end{subfigure}

  \begin{subfigure}[t]{.49\columnwidth}
    \centering
    \includegraphics[width=0.82\textwidth]{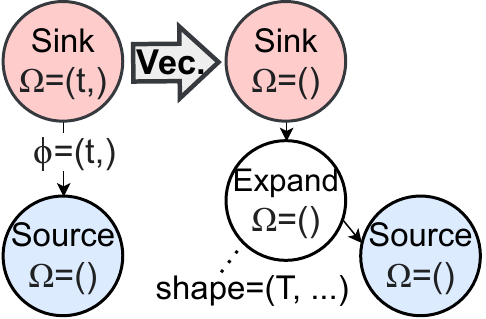}
    \caption{Sink~($t \notin \Omega(\textnormal{source})$)}\label{fig:vec:vec_sink_src_no_dim}
  \end{subfigure}
  \begin{subfigure}[t]{.49\columnwidth}
    \centering
    \includegraphics[width=0.80\textwidth]{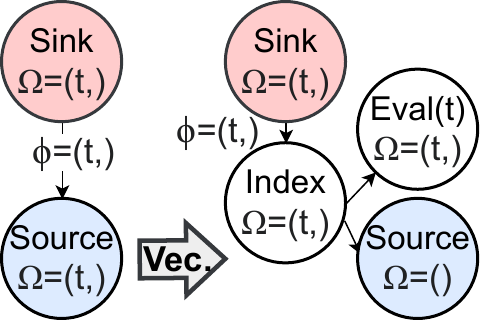}
    \caption{Source}\label{fig:vec:vec_src_sink_has_dim}
  \end{subfigure}
  \caption{How \sys{} updates edges when vectorizing}
  \label{fig:vec:edge_updates}
\end{figure}

\mypar{Handling edges}\sys{} must also update SDG edges for correctness~(see~\F\ref{fig:vec:edge_updates}). If both source and sink are vectorized~(\F\ref{fig:vec:vec_both}) or, if only the source is, and the sink does not have $t$ in its domain~(\F\ref{fig:vec:vec_src_sink_no_dim}), \sys{} drops the $t$ index from the connecting dependence $\phi$. If only the sink is vectorized, the $t$ index is promoted to $0:T$~(\F\ref{fig:vec:vec_sink_src_has_dim}).
If the sink is vectorized, and the source does not originally vary with $t$~(\F\ref{fig:vec:vec_sink_src_no_dim}), the source shape is expanded to the $T$ elements expected by the sink. Finally, if both vary with $t$, but only the source is vectorized~(\F\ref{fig:vec:vec_src_sink_has_dim}), an \textop{IndexSelect} is added to extract the $t$-th spatial element each step~$t$.

\subsection{Tiling}
\label{sec:tiling}

Some operators require too much memory, especially after vectorization. In these cases, \sys{} can \emph{tile} the SDG~(\myc{C} in \F\ref{fig:transformation_pipeline:transforms}). Similar to vectorization, tiling computes a set of operators to tile, applies a tiling rule to each, and updates the edges. Unlike vectorization, however, tiling adds a new temporal dimension~$n$ to the domain of tiled operators and decomposes a chosen spatial dimension of size $D$ into $N$ tiles of size $Z$, with $N=D//Z$.

\mypar{Computing the tiled operator set}
Since reductions (\eg \textop{Sum} and \textop{MatMul}'s contracting dimension) eliminate specific input spatial dimensions, they are natural starting points for tiling. 
\sys{} picks one of the reduced spatial dimensions, $D$, to tile on, preferring dimensions which are eventually introduced by temporal indexing operations. This is because tiling on these enables
online garbage collection during scheduling~(\S\ref{sec:mem_sched}) and thus implicitly enables advanced execution strategies, such as \emph{gradient accumulation}~\cite{huang2019gpipe}.

Next, \sys{} recursively traverses operator dependencies, adding them to the tiled operator set until it encounters one of the following stopping conditions:
\begin{myitemize}
  \item a dependence expression~$\phi$ that creates the spatial dimension being tiled. In this case, \sys{} modifies the expression to instead temporally access the $n$-th tile of size $Z$;
  \item an operator with a memory demand below a user-defined threshold. Here, \sys{} inserts a \textop{Slice} to spatially slice the $n$-th tile of size $Z$ from the non-tiled dependency;
  \item an \textop{Expand} that expands the dimension being tiled. \sys{} modifies the \textop{Expand} to expand to $Z$ instead of $D$.
\end{myitemize}

\noindent
A tiling rule is then applied to each operator in the tiled operator set. These are essentially the opposite of the corresponding vectorization rules.
To finish the tiling, \sys{} performs an aggregation over all $K$ tiles (\eg the white sum of \myc{C} in \F\ref{fig:transformation_pipeline:transforms}), restoring the original reduction.

\begin{figure}[t]
  \begin{subfigure}[t]{.36\columnwidth}
    \centering
    \includegraphics[width=\textwidth, trim={0 1 0 0}, clip]{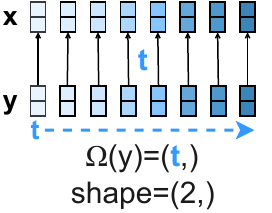}
    \caption{Incremental}\label{fig:rt_transforms:incremental}
  \end{subfigure}
  \hfill
  \begin{subfigure}[t]{.30\columnwidth}
    \centering
    \includegraphics[width=\textwidth, trim={0 1 0 0}, clip]{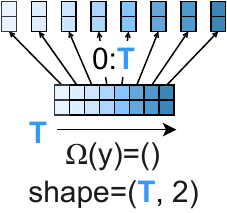}
    \caption{Vectorized}\label{fig:rt_transforms:vectorized}
  \end{subfigure}
  \hfill
  \begin{subfigure}[t]{.305\columnwidth}
    \centering
    \includegraphics[width=\textwidth, trim={0 1 0 0}, clip]{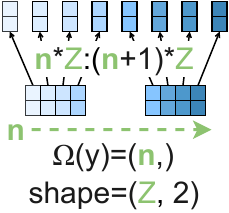}
    \caption{Tiled}\label{fig:rt_transforms:tiled}
  \end{subfigure}
  \caption{Equivalent execution strategies for \boldmath{$y[t]\!=\!-x[t]$} as RTs \textnormal{(Example assumes {\unboldmath $x$} is not also vectorized or tiled.)}}
  \label{fig:rt_transforms}
\end{figure}

\F\ref{fig:rt_transforms} visualizes vectorization and tiling as RTs.
Initially defined by a user \emph{incrementally}~(\F\ref{fig:rt_transforms:incremental}), the identity dependency allows $y[t]$ to execute immediately after $x[t]$. \sys{} can also vectorize $y$~(\F\ref{fig:rt_transforms:vectorized}), which requires that $y$ wait for all $x[0:T]$ values, increasing memory usage. \sys{} can then tile the freshly vectorized dimension~(\F\ref{fig:rt_transforms:tiled}). Note how $y[n]$ now accesses a tile of size $Z=4$ using the dependence expression $\phi=[n \cdot Z\!:\!(n\!+\!1) \cdot Z]$. Fully tiling on the same spatial dimension brings the SDG back to~\F\ref{fig:rt_transforms:incremental}.

\mypar{Static tiling}
In general, dynamic dependencies with changing shapes prevent code-generation. In these cases, \sys{} tiles them into \emph{statically-sized tiles} by adding minimal padding.

\begin{figure}[t]
  \begin{subfigure}[t]{.3225\columnwidth}
    \centering
    \includegraphics[width=\textwidth, trim={0 1 0 1}, clip]{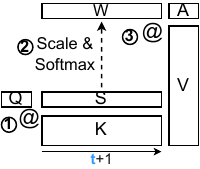}
    \caption{Torch~\cite{paszke2019pytorch}}\label{fig:attention_tiling:torch}
  \end{subfigure}
  \hfill
  \begin{subfigure}[t]{.3225\columnwidth}
    \centering
    \includegraphics[width=\textwidth, trim={0 1 0 1}, clip]{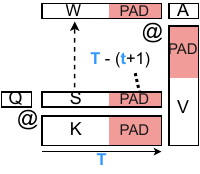}
    \caption{JAX~\cite{frostig2018compiling}}\label{fig:attention_tiling:jax}
  \end{subfigure}
  \hfill
  \begin{subfigure}[t]{.325\columnwidth}
    \centering
    \includegraphics[width=\textwidth, trim={0 2 1 1}, clip]{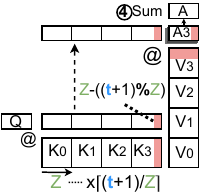}
    \caption{Tempo}\label{fig:attention_tiling:tempo}
  \end{subfigure}
  \caption{\textbf{Attention in DL systems} \textnormal{(Torch computes attention eagerly. JAX optimizes it, but introduces large padding. Tempo tiles it into static tiles, enabling code-generation with minimal padding.)}}
  \label{fig:attention_tiling}
\end{figure}

\F\ref{fig:attention_tiling} shows this for causal attention (\Eq\ref{eq:attention}), which uses $K[t]=k[0\!:\!t\!+\!1]$ to obtain dynamically shaped $K$ and $V$~tensors. As discussed in \S\ref{sec:dataflow}, Torch~\cite{paszke2019pytorch} eagerly executes attention despite the dynamic shapes (\F\ref{fig:attention_tiling:torch}), which prevents optimizations; JAX~\cite{frostig2018compiling} requires padding to a static upper bound~$T$ to enable whole-program optimization, which adds overhead that grows with $T$ (\F\ref{fig:attention_tiling:jax}).

In contrast, \sys{} tiles attention,
trading a dynamic computation for a dynamic number of static tiles~(\F\ref{fig:attention_tiling:tempo}).
Tiling starts from $W \cdot V$, which contracts the dynamic dimension, and continues through the softmax, down to $K$ and $V$. 
This creates $\lceil (t\!+\!1)/Z \rceil$ tiles of size $Z$ for each tensor.

\sys{} uses $K[t,n] = k[n\!\cdot\!Z\!:\!\textrm{min}((n\!+\!1)\!\cdot\!Z, t\!+\!1)]$ to construct each tile, which may not fill $Z$~timesteps due to the use of \texttt{min}.
Thus, it is necessary to pad the remaining amount: $Z-(t\!+\!1)\!\mod\!Z$. 
\sys{} also appropriately masks this padding,
using a $0$ mask for \textop{MatMul}s and \textop{Sum}s, and a $-\infty$ mask for \textop{MaxOp}.
To obtain the final attention output, \sys{} then sums together each masked attention tile.

Padding and masking overhead is minimal, as they are only applied to the last tile. Furthermore, \sys{} can often optimize them away by hinting to the runtime that buffers for $K$ and $V$ tiles should be pre-allocated with the padded size and pre-filled with the mask value (see~\S\ref{sec:runtime}).

\subsection{Operator fusion}
\label{sec:dataflow_fusion}

An SDG often contains static \emph{islands} in which all operators have the same domain and only identity edges between them. Within these islands, the SDG directly passes the outputs of one operator to the inputs of another. The \emph{fusion transformation} identifies and \emph{fuses}~(\myc{D} in \F\ref{fig:transformation_pipeline:transforms}) these islands into a single \textop{Dataflow} operator, routing edges as needed. 

\sys{} initially assigns each operator to its own island and iteratively merges islands with the same domain and no inter-dependencies.
Dynamic operations (\eg \textop{RNG}, \textop{UDF}, \textop{Merge}) are excluded from fusion, because they cannot be compiled statically. Fusion enables \sys{} to leverage existing DL code-generators~\cite{frostig2018compiling, ansel2024pytorch, tillet2019triton} to generate kernels. This lowers dispatching overheads, minimizes the number of tensors \sys{} must manage, and speeds-up scheduling.



\section{Scheduling SDGs with the Polyhedral Model}
\label{sec:analysis}

\begin{figure}[t]
  \centering
  \includegraphics[width=\columnwidth, trim={1 7 1 2}, clip]{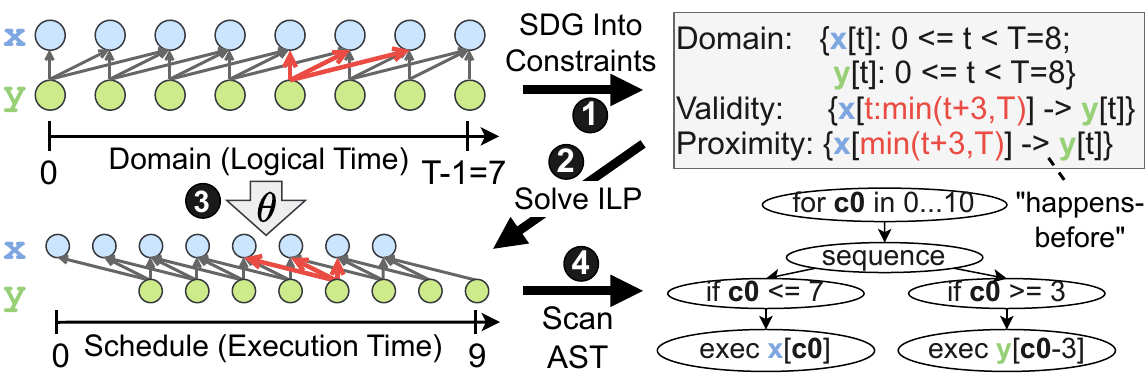}
  \caption{How \sys{} schedules $y[t] = f(x[t\!:\!min(t\!+\!3, T)])$}  
  \label{fig:scheduling_viz}
\end{figure}

RTs are declarative and thus require scheduling to find an execution order. Due to their dynamic dependencies, existing dataflow scheduling techniques~\cite{abadi2016tensorflow,moritz2018ray}, however, are unsuitable. \sys{} uses a \emph{polyhedral model}~\cite{bondhugula2008pluto,verdoolaege2010isl} to address this challenge. Scheduling occurs in two rounds: \sys first finds a schedule for tensor operations~(\S\ref{sec:polyhedral_model}), and then uses it to schedule memory management operations~(\S\ref{sec:mem_sched}).

\subsection{Polyhedral execution scheduling}
\label{sec:polyhedral_model}


The polyhedral model~\cite{verdoolaege2010isl, bondhugula2008pluto} has its origins in the scheduling of uniform recurrence equations~\cite{karp1967organization}, but has since evolved to support more complex computation~\cite{benabderrahmane2010polyhedral}. Today, it is used to optimize localized loop nests (kernels) in imperative programs~\cite{grosser2012polly, vasilache2018tensor}, as using it to model arbitrary programs remains challenging~\cite{verdoolaege2012polyhedral}. Since RTs are inspired by recurrence equations, \sys{} can apply the polyhedral model for whole-program scheduling. 



Scheduling in a polyhedral model finds a \emph{schedule function}~$\theta$ that maps each $n$-dimensional timestep of an operation to an \emph{execution time} in~$\mathbb{Z}^{s}$, $s \geq n$, and where the lexicographic order\footnote{Lexicographic order compares tuples left to right: $(1,2) \prec (1,5) \prec (2,1)$.} ($\prec$) determines what operation executes first. A \emph{valid} schedule must respect all dependencies, \ie  for $y[t]\!=\!f(x[\phi(t)])$ then $\forall t \in \Omega(y) . \theta(x[\phi(t)]) \prec \theta(y[t])$. 
Finding a schedule function requires solving an integer linear program~(ILP)~\cite{verdoolaege2010isl}, with an ILP objective which minimizes the sum of \emph{dependence distances} $\theta(y[t]) - \theta(x[\phi(t)])$ constrained by a set of \emph{validity constraints} derived from program dependencies. This yields schedules with low makespan.

\F\ref{fig:scheduling_viz} shows the scheduling approach in \sys. SDGs already represent tensor domains, dependencies and conditions explicitly using symbolic expressions, which can be translated easily into polyhedral validity constraints (\ding{202} in \F\ref{fig:scheduling_viz}). In addition, \sys{} creates \emph{proximity constraints}, which guide the scheduler towards schedules with higher temporal locality and smaller tensor lifetimes. \sys{} uses the \emph{Pluto} scheduler~\cite{bondhugula2008pluto} through the \emph{isl} library~\cite{verdoolaege2010isl} (\ding{203}). 
In \F\ref{fig:scheduling_viz}, the schedule found delays $y$'s execution by 3 steps.

Since \sys fuses large regions of the SDG, the number of operations is reduced, making the ILP tractable even for large programs. 
\sys{} stores the resulting schedule function~$\theta$ (\ding{204}) for scheduling memory management operations, as discussed next. We describe how \sys scans the schedule to produce an imperative program AST (\ding{205}) in \S\ref{sec:ast_build}.

\subsection{Memory management}
\label{sec:mem_sched}

\begin{figure}[t]
  \centering
  \begin{subfigure}[t]{0.38\columnwidth}
    \centering
    \includegraphics[width=\columnwidth, trim={0 1 1 0}, clip]{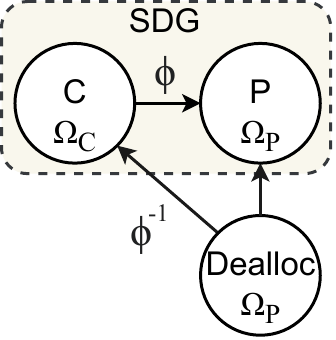}
    \caption{Deallocation}\label{fig:mem_man:dealloc}
  \end{subfigure}%
  \hfill
  \begin{subfigure}[t]{0.58\columnwidth}
    \centering
    \includegraphics[width=\columnwidth, trim={0 1 1 0}, clip]{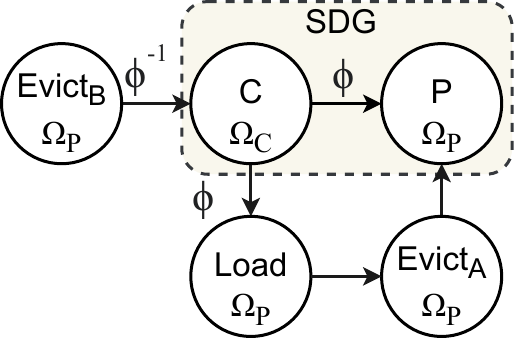}
    \caption{Swap}\label{fig:mem_man:swap}
  \end{subfigure}%
  \caption{SDG memory management augmentations}
  \label{fig:mem_man_augment}
\end{figure}

Memory management is a key concern in DL systems~\cite{peng2020capuchin,rhu2016vdnn,wang2018superneurons}. Using the polyhedral model, \sys{} automates memory management concerns: (i)~deallocations, (ii)~GPU/CPU memory swapping, and (iii)~buffer donations. 

The key idea is to \emph{augment} the SDG with memory management operations (and constraining dependencies), which enable \sys{} to reuse the polyhedral scheduler to find a memory management schedule.

\mypar{Scheduling deallocations} 
\F\ref{fig:mem_man:dealloc} shows how \sys{} augments the SDG with deallocations for the tensor produced by $P$ and consumed by $C$. Since $P$ produces a tensor for each timestep in its domain, $\Omega_P$, the deallocation $\mathrm{Dealloc}$ shares this same domain. \sys{} interprets the execution of $\mathrm{Dealloc}[t]$ as the time to deallocate the tensor $P[t]$.

Deallocation must always happen after production, which \sys{} models with an identity dependence, $\mathrm{Dealloc} \rightarrow P$. 
Then, to also ensure $P[t]$ is deallocated only after the consumer has finished using it, \sys{} adds a dependence from $\mathrm{Dealloc}$ to $C$, with the inverse dependence expression $C$ has on $P$. 
For example, if $\phi\!=\!t-3$, then $\mathrm{Dealloc}[t]$ must wait for $C[t+3]$ to run, before it can deallocate $P[t]$.

\mypar{Scheduling GPU/CPU memory swapping} 
\F\ref{fig:mem_man:swap} shows how \sys{} augments the SDG with evict (GPU-to-CPU) and load (CPU-to-GPU) memory movement operations. 
\sys{} inserts an initial evict operation $\mathrm{Evict_A}$, which evicts the tensor $P[t]$ immediately after production.
It then inserts $\mathrm{Load}$ and $\mathrm{Evict_B}$ operators, to re-load and again evict all timesteps of $P$ needed by $C$ at timestep $t$. All operations have domain $\Omega_P$ and \sys{} interprets the execution of $\mathrm{Evict_{A/B}}[t]$ and $\mathrm{Load}[t]$ as the time to evict or load $P[t]$, respectively.

A tensor can only be evicted after it has been produced, and loaded after it has been evicted. \sys{} models this with the identity dependencies, $\mathrm{Load} \rightarrow \mathrm{Evict_A} \rightarrow P$.
To guarantee that the timesteps of $P$ needed by $C[t]$ are available in-memory when $C[t]$ executes, \sys{} adds $C \xrightarrow{\smash{\phi}} \mathrm{Load}$.
Finally, similarly to deallocations, $\mathrm{Evict_B}$ must execute only after the last use of $P[t]$ by $C$, so \sys{} adds the inverse of the consumer's dependence from $\mathrm{Evict_B}$ to $C$.

By placing these dependencies, \sys{} introduces constraints on when the memory management operations can execute, which guarantees that the resulting schedule is valid and efficient.
To prevent thrashing~\cite{denning1968working}, \sys{} does not currently swap frequently accessed tensors, such as DNN parameters or K/V caches in attention models.
At runtime, \sys{} performs such swap operations \emph{asynchronously}, through a shared \emph{page-locked} host-memory manager, which pre-allocates page-locked buffers for each swapped RT.

\mypar{Finding buffer donations} Given the SDG's stateless nature, it is also important to detect when a tensor produced by one operation can be \emph{donated} to be reused by another operation. This avoids expensive re-allocations and enables in-place mutation of, \eg DNN parameters.

After a tensor is donated, it is invalidated and can no longer be used by other operations.
Thus, an operator $O_d$ can only donate its output tensor to one of its dependents, called the \emph{donation receiver} $O_r$, if for every timestep $t$ in its domain $\Omega(O_d)$, $O_r$ is the last user of $O_d[t]$. This requires that it is scheduled \emph{after} all other competing dependents $O_c$ according to the schedule function~$\theta$. Formally, \sys donates $O_d$ to $O_r$ if and only if
\begin{align*}
\forall t \in \Omega(O_d) .\ \forall O_c \in \mathcal{D}(O_d) \setminus \{O_r\} .\ \theta\left(\phi^{-1}_{c}(t)\right) \prec \theta\left(\phi^{-1}_{r}(t)\right)
\end{align*}
where $\mathcal{D}(O_d)$ is the set of dependents of $O_d$, $\phi^{-1}_{c}$ and $\phi^{-1}_{r}$ are the inverse dependence expressions from $O_c$ and $O_r$ to $O_d$ respectively, and $\setminus$ is the set difference operator. 

\subsection{Generating imperative programs}
\label{sec:ast_build}

\sys{} must obtain an imperative program for the user's declarative RT computation, but the schedule function~$\theta$ produced by the polyhedral scheduler~\cite{bondhugula2008pluto} is not directly executable. \sys{} transforms it into an abstract syntax tree~(AST) by using the polyhedral library~\cite{verdoolaege2010isl} to scan~\cite{grosser2015polyhedral} it. The AST contains the following types of statements:
\begin{myitemize}
  \item \textbf{sequence}: a block of AST nodes executed in order;
  \item \textbf{if}: a conditional with then and else sub-ASTs;
  \item \textbf{loop}: a generic loop with an iteration counter, start value, condition, step and body sub-AST;
  \item \textbf{execute}: executes operation~$O[t]$, where $t$ is given by a mapping from loop counters; and
  \item \textbf{deallocate/load/evict}: executes the memory instruction for timestep~$t$ (or multiple steps at once) given by a mapping from loop counters.
\end{myitemize}

To generate efficient ASTs, \sys applies further optimizations: (1)~it \emph{partially unrolls} loops to reduce the number of nested conditionals; (2)~it \emph{promotes} static loops that contain only memory management operations into batched instructions by mapping the loop's range to a slice of points; and (3)~it finds and removes redundant loads/evicts by removing (i)~any evict followed by a load or deallocation for the same tensor; and (ii)~loads for already in-memory tensors. 

By interpreting the AST at runtime, users are able to step through the program with high granularity. This allows for simpler debugging and integration with existing DL libraries.



\section{Execution Runtime Implementation}
\label{sec:runtime}

The core of \sys{} is written in $\sim$33K Python LoC. Its runtime (rightmost block in \F\ref{fig:overview}) is responsible for executing the program given an SDG, imperative AST, and DL backend.  

The runtime executor is simple: it traverses the schedule AST from the root, keeping track of loop counters as they are introduced, evaluates if conditions and follows each execute/deallocate/swap instruction in order. We highlight four key components of the runtime:

\myparr{DL execution backends} implement a thin interface, which has methods for (i)~allocating tensors, (ii)~moving tensors between GPU/CPU, (iii)~translating each \sys{} tensor operation to the backend equivalent, and (iv)~a code-generation function that returns a kernel when given a fused \textop{Dataflow}. \sys{} currently offers Torch~\cite{paszke2019pytorch}~(without buffer donation support) and JAX~\cite{frostig2018compiling} DL backends, easing integration with existing projects.

\myparr{Kernel launchers} manage the inputs and outputs of kernel invocations. When the executor encounters an execute instruction, it invokes the corresponding kernel launcher with the loop counter state. Using the loop counters, kernel launchers evaluate input dependence expressions and use the result to index tensor stores for the inputs to the kernel. They then execute the kernel and store the outputs back in tensor stores at the current timestep $t$. 

\myparr{Tensor stores} are how \sys{} stores the timesteps of RTs at runtime. Tensor stores are always written into timestep-by-timestep, but are accessed using various patterns, which requires tailored storage strategies for best performance. 
\sys{} assigns each RT to one of three storage methods:

  \begin{myitemize}
    \item If the RT is point accessed (see~\F\ref{fig:access_patterns:state_passing}), \sys{} stores it in a \emph{point store}, which maps the point directly to a physical tensor. This makes point insertions and reads fast, at the cost of requiring tensor stacking for slice reads;
    \item If the RT is block accessed (see~\F\ref{fig:access_patterns:block}) \sys{} stores it in a \emph{block store}, which maps the access to a block-sized pre-allocated buffer. It offers fast slice reads, but writes require in-place mutation. Causally and anti-causally accessed RTs also use this store (with buffer size $T$);
    \item If the RT is window accessed (see~\F\ref{fig:access_patterns:fwd_window}), \sys{} stores it in a \emph{window store}, which maps the access to a pre-allocated circular buffer twice the size of the window. \sys{} writes to two mirrored locations, ensuring that a contiguous read window is always ready.
  \end{myitemize}

\begin{figure}[t]
  \centering
  \includegraphics[width=0.95\columnwidth, trim={0 1 1 0}, clip]{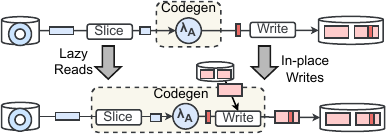}
  \caption{Kernel wrappers \textnormal{(\sys{} avoids copies by co-designing the storage and code-generation of kernels.)}}\label{fig:kernel_wrappers}
\end{figure}

 \myparr{Kernel wrappers} inject storage related operations into kernels prior to code-generation~(see~\F\ref{fig:kernel_wrappers}). This enables efficient interaction with block or window stored RTs. Normally, a kernel first produces an output tensor, which is then copied into a pre-allocated buffer at index $t$.  The \emph{in-place writes} wrapper eliminates such copies by enabling the kernel to write directly to the preallocated buffer.  Conversely, some DL backends (\eg JAX) perform copies when slicing tensors. The \emph{lazy reads} wrapper optimizes this by slicing the backing buffer in the code-generated region.



\section{Evaluation}
\label{sec:eval}

We evaluate \sys{} to answer the following questions: do \sys{}'s optimizations lead to more efficient computation compared to existing approaches? 
does \sys{}'s tiling and memory management enable scaling on a single GPU? 
can \sys{} effectively support algorithm-specific execution and memory-management schedules? 
does \sys{}'s compilation scale to large model sizes? 

\subsection{Methodology}
\label{sec:eval:methodology}

We evaluate \sys on LLM decoding and RL training.

\mypar{LLM decoding} We focus on decoding, because it often dominates LLM inference latency~\cite{agrawal2024taming}. We implement a Llama-3.2-3B~\cite{dubey2024llama} decoding loop (\F\ref{fig:transformer_decoding}) in \sys{}, and compare it against end-to-end JIT-compiled JAX~\cite{frostig2018compiling} (a graph-based system), and Torch~\cite{paszke2019pytorch}\footnote{We use Meta's open-source inference implementation from \\\url{https://github.com/meta-llama/llama3}.} (an eager-execution system). For the JAX baseline, we pad the minimal amount possible (\ie up to the sequence length being decoded). We compare the performance as we modify batch size, sequence length, and the attention pattern from causal~\cite{vaswani2017attention} to windowed~\cite{zaheer2020big}. All systems use the same tokenizer, sampling method, and prompts.

\mypar{RL training} We experiment with (a)~PPO~\cite{schulman2017proximal}, (b)~REINFORCE~\cite{williams1992simple} and (c)~REINFORCE with $n$-step returns~\cite{sutton1988learning} (essentially A2C~\cite{mnih2016asynchronous}).
We compare \sys{} against 5 RL frameworks with different focus: (i)~performance-focused frameworks (SampleFactory~v2.1.1~\cite{Petrenko2020SampleFE}, RLGames~v1.6.1~\cite{rl-games2022}), (ii)~single-file algorithm implementations (CleanRL~commit~e648ee2~\cite{huang2022cleanrl}), and (iii)~scalable RL frameworks (Ray RLlib~v2.5.0~\cite{liang2018rllib}). To focus on framework overheads, we employ a test RL environment~\cite{freeman2021brax, makoviychuk2021isaac} with short simulation-step time for all experiments. It is GPU-accelerated except for RLlib, which lacks such support. 




All experiments use the same hardware and software configuration. We use a server with an AMD EPYC 7402P 24-core CPU, 384\unit{GB} of DDR4 RAM (3200\unit{MT/s}) and an NVIDIA RTX A6000 GPU (48\unit{GB} of GDDR6 RAM, PCIe Gen4 $\times$16 at 32\unit{GB/s}). We run Ubuntu~v22.04 with Linux kernel~v5.15, CUDA~v12.8, PyTorch~v2.7.1, and JAX~v0.6.2. We discard the first iteration of each experiment to avoid warmup effects.

\begin{figure*}[t]
  \centering
  \begin{subfigure}[t]{0.356\textwidth}
    \centering
    \includegraphics[width=\linewidth, trim={2 5 3 1}, clip]{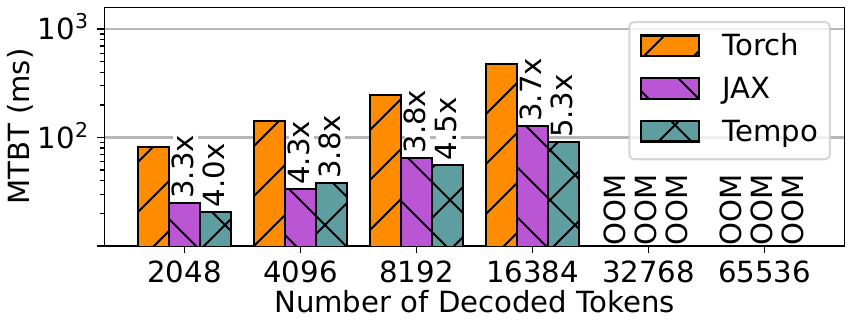}
    \caption{Causal attention~\cite{vaswani2017attention} at batch size 16}
    \label{fig:llama:causal_attention_16}
  \end{subfigure}%
  \hfill
  \begin{subfigure}[t]{0.318\textwidth}
    \centering
    \includegraphics[width=\linewidth, trim={45 5 3 1}, clip]{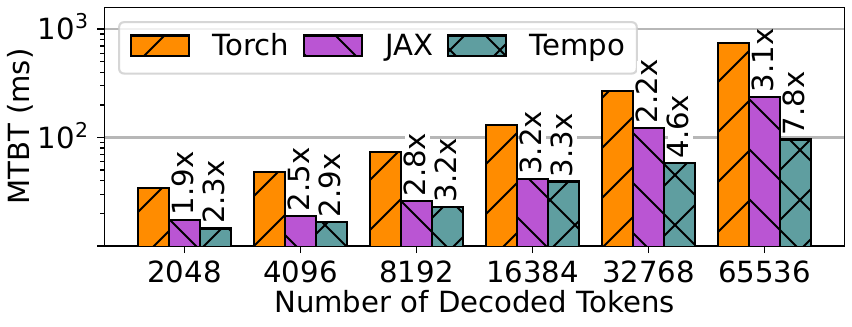}
    \caption{Causal attention~\cite{vaswani2017attention} at batch size 4}
    \label{fig:llama:causal_attention_4}
  \end{subfigure}
  \hfill
  \begin{subfigure}[t]{0.318\textwidth}
    \centering
    \includegraphics[width=\linewidth, trim={45 5 3 1}, clip]{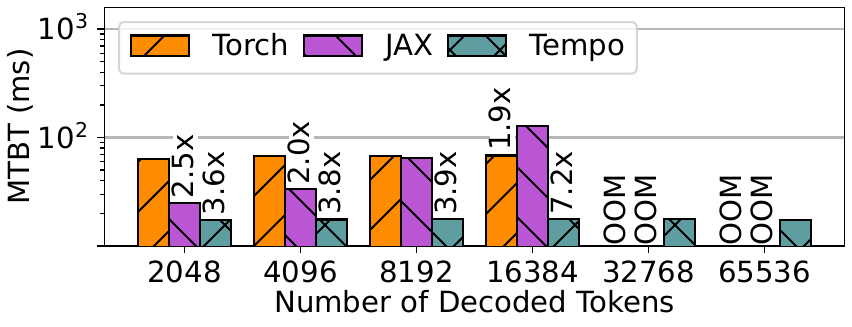}
    \caption{Window attention~\cite{zaheer2020big} at batch size 16}
    \label{fig:llama:window_attention_16}
  \end{subfigure}%
  \caption{Llama-3.2-3B mean time between tokens (MTBT) with different attention mechanisms and batch sizes}
  \label{fig:llama:throughput}
\end{figure*}

\subsection{LLM decoding}
\label{sec:eval:decoding}

We begin our evaluation with LLama-3.2-3B~\cite{dubey2024llama} token-by-token decoding. For each system, we fix the batch size to $16$ and measure the latency in terms of mean time between tokens (MTBT) when decoding sequences of length~$T\!=\!2,048$ to $T\!=\!65,536$ tokens with standard causal attention. \F\ref{fig:llama:causal_attention_16} shows the results, where the factors indicate speedups over the slowest system. 
In general, as we increase the sequence length, the MTBT of all systems increases. This is expected as the dynamic $[0\!:\!t\!+\!1]$ dependency of causal attention increases the amount of work with each added timestep.

Torch is consistently the slowest system, and its MTBT grows the fastest. This is because, as an eager-execution system, Torch does not make use of compilation-based optimizations. For $T\!<\!4,096$, \sys{} and JAX behave similarly,
as both process the entire sequence as a single padded tensor. 
For $T\!\geq\!4,096$, however, JAX continues to add padding, while \sys{} begins tiling attention (\S\ref{sec:tiling}). 
This causes an initial slow-down at $\!T=\!4,096$ as \sys{}'s tiling reduces parallelism, but a growing speedup for $T\!>\!4,096$, due to reduced padding overhead.
\sys{} is up to 43\% faster at $T\!=\!16,384$, and while this gap should grow with $T$, all systems run out of GPU memory at $T\!=\!16,384$ due to K/V cache pressure.

\mypar{Decoding long sequences}
To scale to longer sequences, we lower the batch size to $4$ and repeat the experiment. \F\ref{fig:llama:causal_attention_4} shows the results. Similar behavior is observed, with 16,384 now serving as the point at which \sys{}'s tiling reduces parallelism and JAX starts accumulating padding overhead. However, the longer sequence lengths cause more drastic slowdown for JAX, as the amount of padding it processes grows linearly with $T$. At $T\!=\!32,768$, \sys{} is 2$\times$ faster than JAX, and that grows to 2.5$\times$ faster at $T\!=\!65,536$.

\begin{figure}[t]
  \centering
  \includegraphics[width=\columnwidth, trim={4 8 6 7}, clip]{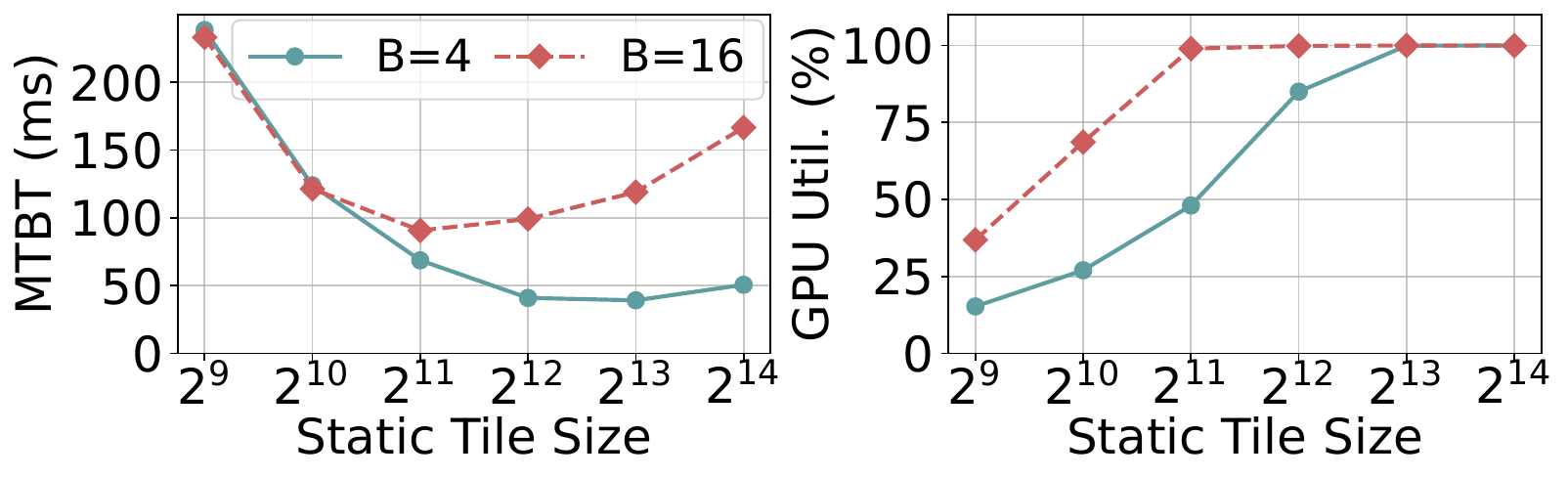}
  \caption{Effect of \sys{}'s tile size on Llama-3.2-3B}\label{fig:llama:block_size}
\end{figure}

\mypar{Static tile size} These results can be better understood by studying how \sys{}’s tiling strategy (\S\ref{sec:tiling}) impacts decoding performance. The tile size used to decompose causal attention into statically-sized tiles is a key parameter that controls the trade-off between parallelism and padding overhead.

\F\ref{fig:llama:block_size} shows the tile size affects latency for a fixed decode length of $T\!=\!16,384$ tokens and varying batch size $B$. At $B\!=\!4$, increasing the tile size up to $Z\!=\!8,192$ progressively lowers MTBT despite adding large padding, because the GPU has spare capacity to process this padding.
At $BS\!=\!16$, however, MTBT decreases until the GPU utilization is maximized with $Z\!=\!2,048$, at which point it begins increasing again, yielding a \emph{u-shaped} latency curve. This is because padding adds overhead when there is no spare GPU capacity. For medium to long sequence lengths, JAX is essentially in the low spare GPU capacity regime, and thus it is affected by padding overheads.

\mypar{Windowed attention} Another method for scaling LLMs to longer sequences is to switch to a windowed attention pattern. \F\ref{fig:llama:window_attention_16} shows the results for Llama-3.2-3B with windowed attention at $B\!=\!16$. \sys{} and Torch achieve near constant MTBTs due to their support for dynamic shapes. Torch is eager however, while \sys{} enjoys whole-program compilation, making \sys{} up to 3.9$\times$ faster. JAX retains the exact same MTBT profile as for causal attention (\F\ref{fig:llama:causal_attention_16}) since it models dynamic dependencies using masking over a spatial dimension. At $T\!=\!16,384$ \sys{} is up to $7\times$ faster. Furthermore, \sys{} is the only system that adapts memory management to window dynamic dependencies. \sys{} uses a circular store for K/Vs and infers when old K/Vs can be deallocated, scaling to at least 4$\times$ longer sequences.

\begin{figure}[t]
  \centering
  \includegraphics[width=0.975\columnwidth, trim={1 8 1 1}, clip]{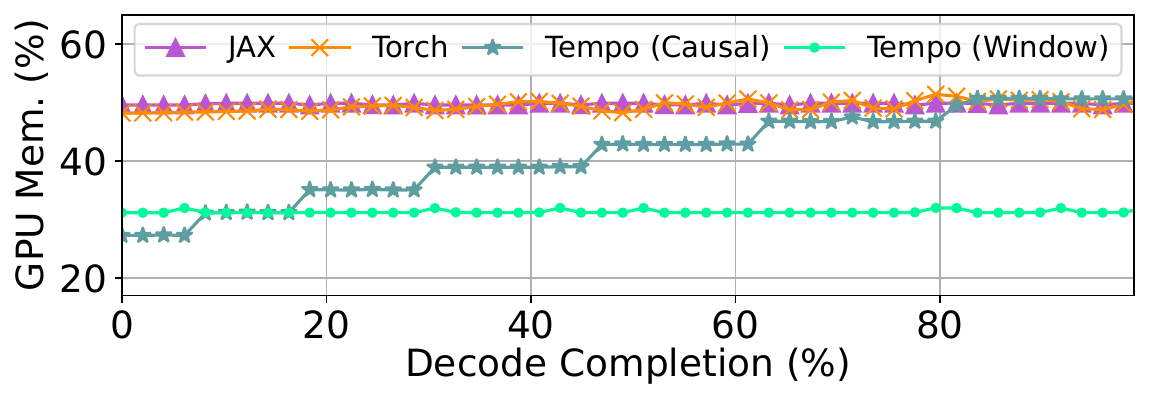}
  \caption{Runtime GPU memory use for Llama-3.2-3B decoding}\label{fig:llama:memory_usage}
\end{figure}

These conclusions are further confirmed by examining the runtime GPU memory usage (\F\ref{fig:llama:memory_usage}). JAX and Torch retain the same memory usage behavior between causal and windowed attention, \ie they preallocate large buffers up-front and never deallocate memory. In contrast, \sys{} effectively manages memory according to algorithm-specific dynamic dependencies.
For causal attention, \sys{}'s memory usage grows only as \sys{} requires more static tiles to be processed, yielding a stepped usage pattern; for windowed attention, \sys{}'s circular buffer tensor storage uses a small static allocation for all sequence lengths.

\begin{figure*}[t]
  \centering
  \includegraphics[width=0.9\textwidth, trim = {4cm 17cm 4cm 0.25cm}, clip]{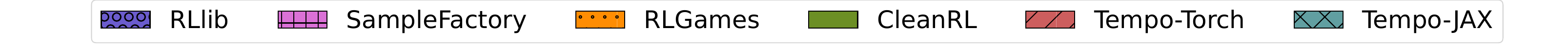}

  \begin{subfigure}[t]{0.273\textwidth}
    \centering
    \includegraphics[width=\columnwidth, trim = {0.1cm 1.4cm 44.75cm 2.85cm}, clip]{cr_figures/eval/rl/small_to_med_scale_no_cleanrl_c}
    \caption{Simulation length}\label{fig:small-scale:episodes}
  \end{subfigure}%
  ~
  \begin{subfigure}[t]{0.229\textwidth}
    \centering
    \includegraphics[width=\columnwidth, trim = {17.8cm 1.4cm 29.90cm 2.85cm}, clip]{cr_figures/eval/rl/small_to_med_scale_no_cleanrl_c}
    \caption{Parameters per layer}\label{fig:small-scale:parameters}
  \end{subfigure}%
  ~
  \begin{subfigure}[t]{0.229\textwidth}
    \centering
    \includegraphics[width=\columnwidth, trim = {32.6cm 1.4cm 15.05cm 2.85cm}, clip]{cr_figures/eval/rl/small_to_med_scale_no_cleanrl_c}
    \caption{Number of hidden layers}\label{fig:small-scale:hidden_layers}
  \end{subfigure}%
  ~
  \begin{subfigure}[t]{0.229\textwidth}
    \centering
    \includegraphics[width=\columnwidth, trim = {47.5cm 1.4cm 0.16cm 2.85cm}, clip]{cr_figures/eval/rl/small_to_med_scale_no_cleanrl_c}
    \caption{Number of environments}\label{fig:small-scale:envs}
  \end{subfigure}%
  \caption{End-to-end iteration time and GPU utilization for PPO at small-to-medium scale.}
  \label{fig:small-scale}
\end{figure*}

\subsection{Training performance for RL}
\label{sec:eval:rl}


We evaluate the end-to-end RL training performance of \sys{} with PPO, a popular RL algorithm. PPO includes an anti-causal dependency~(\F\ref{fig:access_patterns:anti_causal}) that prevents parallel acting and learning. We start with a default configuration with a simulation length of $250$, 512~environments, 64~parameters per layer, 2~hidden layers, and a $3\tighttimes4\tighttimes4$ observation shape and then vary each parameter individually.

\mypar{Small-to-medium scale PPO} \F\ref{fig:small-scale} shows the result by reporting iteration time, GPU utilization, and speed-up factors over the slowest system. \sys{} has consistently lower iteration times than all other baselines: it is up to 54$\times$ faster than RLlib, and on average 2.6$\times$ faster than the next fastest system, CleanRL, which is a hand-optimized Torch PPO implementation. This is due to \sys{}'s whole-program compilation: \sys{} deduplicates the forward passes of actor and learner by caching actor activations. In addition, \sys{}'s vectorization of the timestep dimension, efficient scheduling, and code-generation further improve performance.

As simulation length grows~(\F\ref{fig:small-scale:episodes}), all systems slow down linearly, narrowing relative performance differences. Adjusting the number of parameters per layer~(\F\ref{fig:small-scale:parameters}) does not affect relative performance until $256$ parameters, when \sys{} begins tiling the backward pass.
Increasing hidden layers~(\F\ref{fig:small-scale:hidden_layers}) slows down all systems, but particularly \sys{}. This is likely due to Python-level AST interpretation overheads, which become apparent at this smaller scale and could be avoided by using an efficient C++ AST executor. We confirm this by the fact that adding environments~(\F\ref{fig:small-scale:envs}) improves \sys{}'s relative performance, because AST interpretation overheads are amortized.

\sys{}'s Torch backend is consistently slower (by 2$\times$ on average) than the JAX backend. This highlights the importance of pairing \sys{} with a capable code-generator such as XLA~\cite{openxla2024}, with support for Triton~\cite{tillet2019triton} to find optimized kernels, CUDAGraphs~\cite{cudagraphs} to lower dispatch overheads, and buffer donations~(\S\ref{sec:mem_sched}) to avoid allocations.

At this scale, the GPU is generally underutilized by all systems. However, \sys{}'s GPU utilization is on average higher than the other systems, which indicates lower dispatch overheads due to our vectorization and dataflow fusion optimizations. Peak utilization is also high for RLlib, but this is due to CPU/GPU data transfers performed every timestep.

\begin{figure}[t]
  \centering
  \includegraphics[width=0.9955\columnwidth, trim={4 10 6 6}, clip]{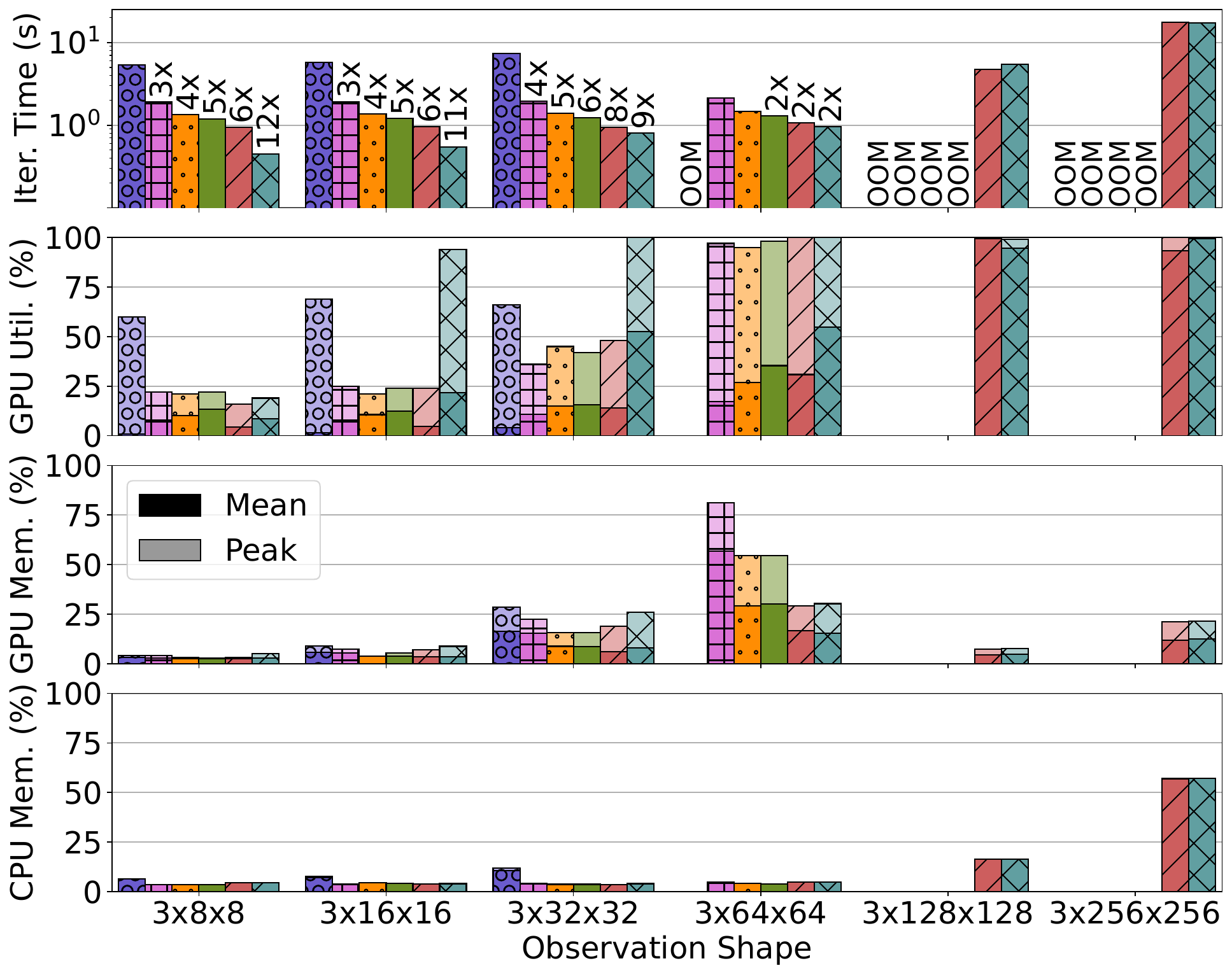}
  \caption{Scaling to large observation sizes with PPO}
  \label{fig:large_scale_range}
\end{figure}

\mypar{Scaling to large observations} Training RL models from image observations is a common use case~\cite{handa2023dextreme, kiran2021deep, yu2021reinforcement} that actor-learner RL frameworks struggle with. Higher image resolutions improve task performance, but require more memory. In this experiment, we halve the number of environments to 256, set the simulation length to 1,000, and show iteration times, GPU utilization and memory, and CPU memory, as we scale observations up to $3\tighttimes256\tighttimes256$.

As \F\ref{fig:large_scale_range} shows, the actor-learner approach stores and learns from all observations at once, resulting in a GPU peak memory usage that exceeds capacity. RLlib enters a fail-retry loop before $3\htighttimes64\htighttimes64$, while the other baselines fail to scale due to out-of-memory~(OOM) errors. In contrast, \sys{} scales to $3\tighttimes256\tighttimes256$ (16$\times$ larger) by balancing GPU/CPU memory usage through tiling and swapping (see~\S\ref{sec:tiling} and \S\ref{sec:mem_sched}), without the need for a replay buffer. 
\sys uses tiling after $3\tighttimes64\tighttimes64$, leading to a slight slow-down, and swapping is used after $3\tighttimes128\tighttimes128$. To scale PPO further, \sys would only require adding more CPU memory.

\begin{figure}[t]
  \centering
  \includegraphics[width=\columnwidth, trim={2 7 2 6}, clip]{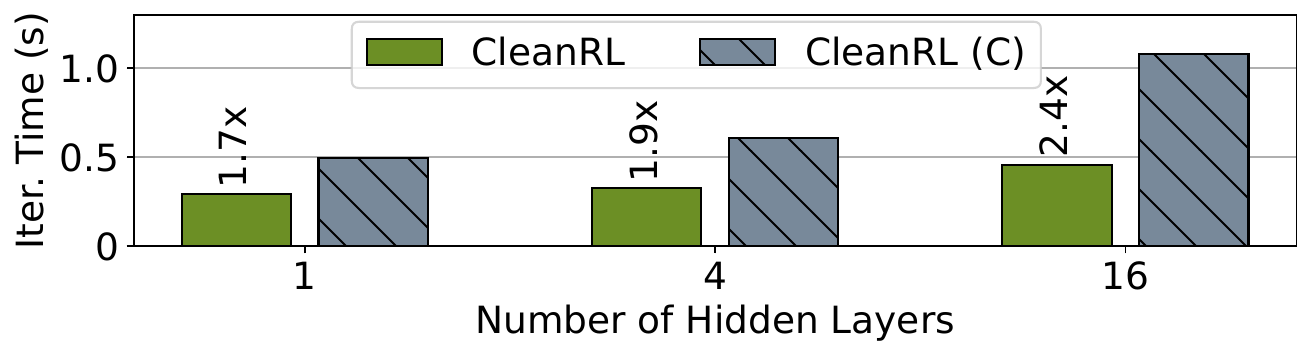}
  \caption{Activation caching without symbolic differentiation}
  \label{fig:cleanrl_vs_cleanrlcache}
\end{figure}

\mypar{Activation caching} We modify CleanRL to also cache and reuse forward pass activations during learning. \F\ref{fig:cleanrl_vs_cleanrlcache} shows that CleanRL's iteration time is on average 2$\times$ slower with caching~(CleanRL~(C)). In Torch, which CleanRL uses, each forward pass creates a corresponding backpropagation graph. The full backward graph concatenates each timestep's graph, causing \emph{graph size explosion}. Thus, Torch must process thousands of small subgraphs, incurring large overhead. In \sys{}, forward pass activations have an explicit timestep dimension, which is automatically vectorized~(\S\ref{sec:vectorization}), enabling an efficient backpropagation graph.


\subsection{Algorithm-specific scheduling}
\label{sec:eval:alg_scheduling}

\begin{figure}
  \centering
  \begin{subfigure}[t]{\columnwidth}
    \centering
    \includegraphics[width=\columnwidth, trim={9 8 6 4}, clip]{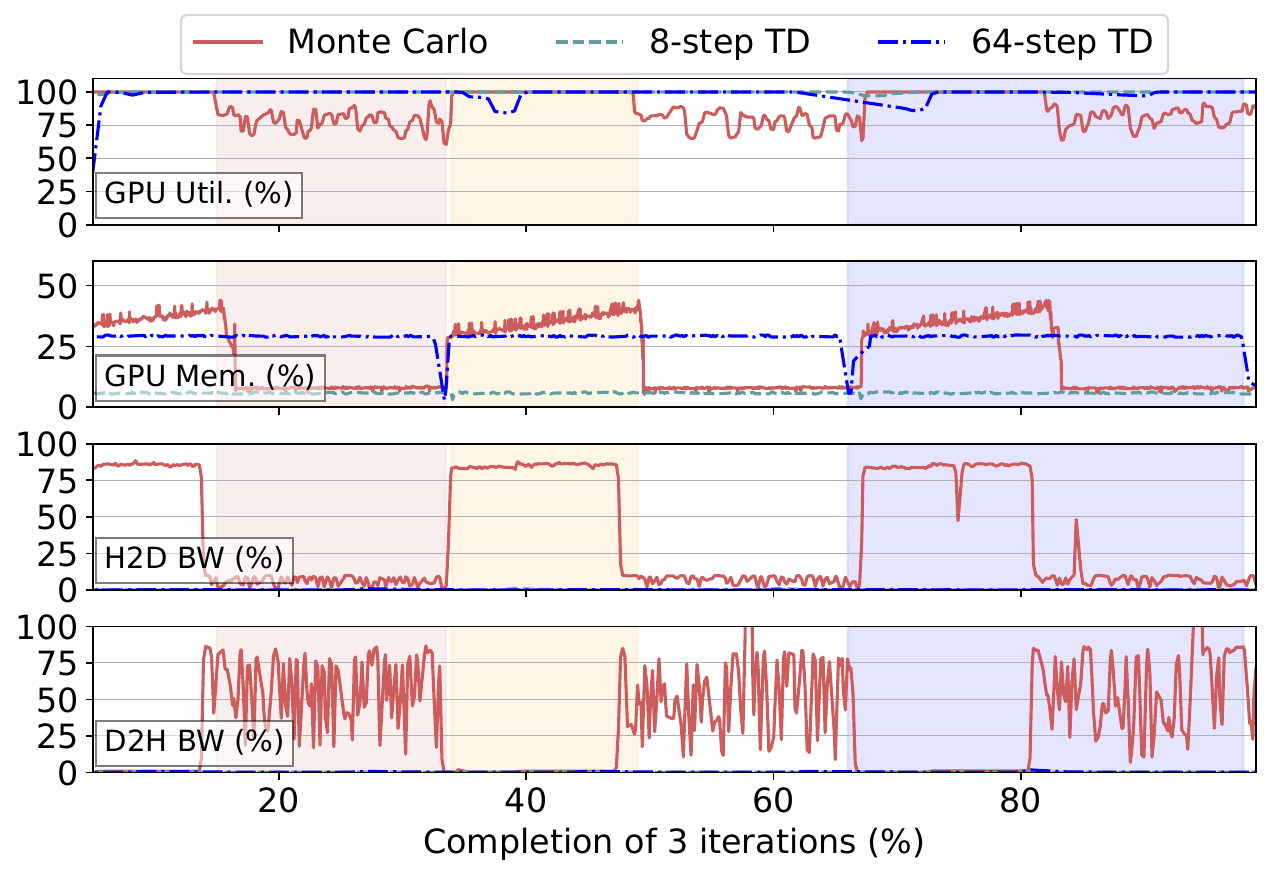}
    \caption{Aligned runtime metrics for different algorithms}
    \label{fig:alg_scheduling:runtime_metrics}
  \end{subfigure}
  \begin{subfigure}[t]{0.95\columnwidth}
    \centering
    \includegraphics[width=\columnwidth, trim={8 8 8 1}, clip]{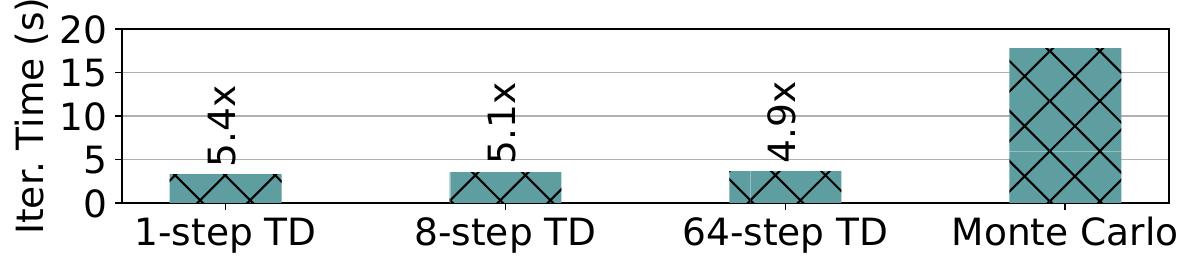}
    \caption{Iteration times for REINFORCE variations}
    \label{fig:alg_scheduling:iter_time}
  \end{subfigure}
  \caption{Algorithm-specific scheduling with \sys{}-JAX}
  \label{fig:alg_scheduling:combined}
\end{figure}

We explore \sys{}'s ability to adapt scheduling to algorithm-specific dynamic dependencies in the large observation setting ($3\tighttimes256\tighttimes256$). We use the two variations of REINFORCE shown in \A\ref{alg:experimentation}, which use: (i)~traditional Monte Carlo~\cite{williams1992simple} returns; and (ii)~$n$-step temporal-difference~(TD) returns~\cite{sutton1988learning}.

\F\ref{fig:alg_scheduling:runtime_metrics} shows how \sys{} executes these two algorithms. Since Monte Carlo (red line) uses an anti-causal access pattern ($r[t:T]$), \sys{}'s scheduler waits until the simulation finishes before learning. During simulation  (red shading), \sys{} continually evicts observations to CPU memory, maintaining low GPU memory usage, but not fully utilizing the GPU. Once finished, learning begins (orange shading), and \sys{} incrementally swaps observations back to GPU memory, learning from them while fully utilizing the GPU.

With $n$-step TD (blue and teal lines), however, \sys{}'s scheduler finds a different strategy: due to the forward window access pattern ($r[t\!:\!min(t\!+\!n, T)]$), after an $n$-step delay, it begins computing gradients \emph{in parallel} with simulation (blue shading). This enables \sys{} to: (i)~fully utilize the GPU; (ii)~only store the last $n$~observations, deallocating older data as early as possible; and (iii)~avoid swapping due to the lower memory demand. Due to the $64$-step delay before learning starts, $64$-step TD still experiences a slight GPU utilization dip, which $8$-step TD does not. 

For smaller $n$, \sys{} parallelizes acting and learning sooner, leading to faster iteration times, as shown in \F\ref{fig:alg_scheduling:iter_time}.
Note that actor-learner frameworks~(see~\S\ref{sec:frameworks}) cannot use such parallel strategies due to coarsening dynamic dependencies between acting and learning to $[0\!:\!T]$.
For either algorithm, \sys{} finds an effective schedule,
allowing for large-scale training on a single GPU.

\subsection{Compilation scalability}
\label{sec:eval:compilation_scalability}

\begin{figure}
  \centering
  \includegraphics[width=\columnwidth, trim={4 7 14 6}, clip]{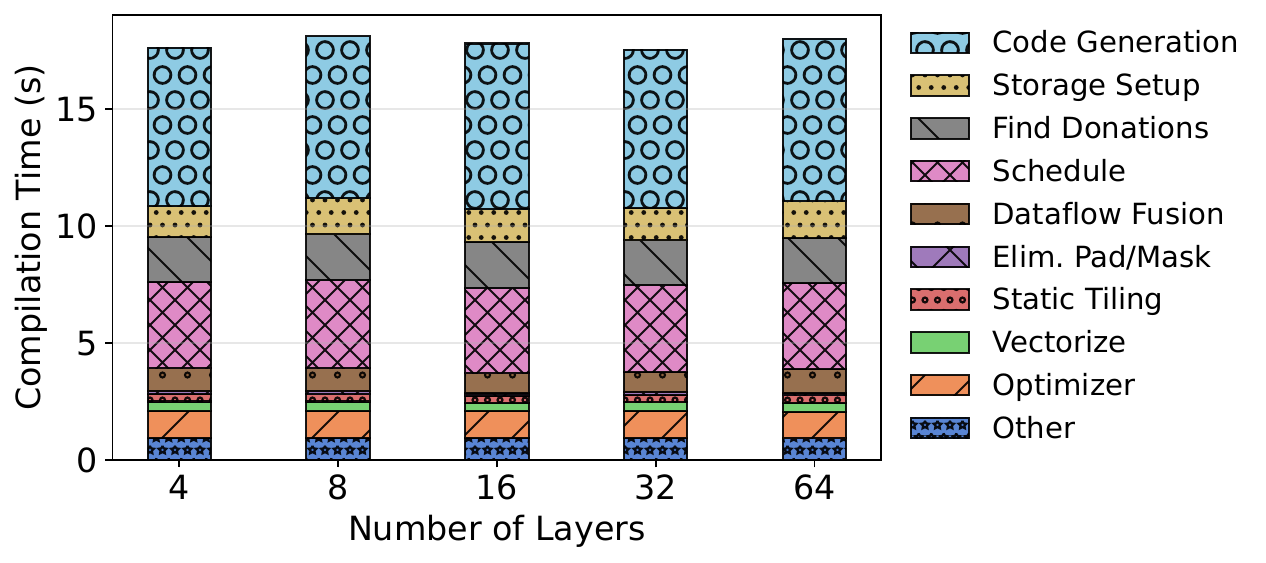}
  \caption{Compilation breakdown for different model sizes}
  \label{fig:compilation_breakdown}
\end{figure}

We evaluate the scalability of \sys{}'s compilation process by varying the number of transformer blocks in the Llama-3.2 architecture.  \F\ref{fig:compilation_breakdown} shows that the compilation time remains near constant as the model size is increased. This is due to \sys{}'s ability to exploit repeated structure in models with a temporal dimension $l$ for layers, creating small SDGs even for large models. Each layer~$l$ depends upon the previous layer~$l\!-\!1$, and the final output is given by layer~$L\!-\!1$.

Compilation times are around 18\unit{s}, of which 7\unit{s} are spent in DL backend code-generation and 1\unit{s} is spent on initial allocations for tensor stores. Beyond this, the majority of the compilation time is taken by polyhedral scheduling and buffer donation analysis. This is because these steps require invocations of the polyhedral library~\cite{verdoolaege2010isl}, which parses constraint strings. Due to \sys{}'s fusion and layer encoding, ILP solving accounts for only 2\unit{s} of the scheduling time.
The remaining compiler passes are fast, because they involve only SDG and symbolic manipulations. Vectorization, tiling and the optimizer each take under 1\unit{s} to complete, despite their fairly complex transformations. Dataflow fusion takes slightly longer, because it performs a large number of validity checks for each fusion. 



\section{Discussion}
\label{sec:discussion}

\mypar{Distribution} While \sys{} is currently a single-GPU system, its temporal tensor dimensions can serve as a foundation for distribution~\cite{silvestre2025rlft}. This requires identifying temporal dimensions that can be mapped to parallel for-loops in the schedule. 
\sys{} can create such temporal dimensions by tiling (\S\ref{sec:tiling}) the SDG across a new ``worker'' dimension. If the starting point is an input-data tensor, this gives \emph{data-parallelism}~\cite{krizhevsky2017imagenet}; in combination with SDG cuts (partitioning the SDG into per-worker subgraphs), input-data tiling also yields \emph{pipeline parallelism}~\cite{huang2019gpipe}; if tiling starts from a weight tensor, we obtain \emph{model-parallelism}~\cite{shoeybi2019megatron}. \sys{} can then assign schedule partitions to workers, which perform reductions as collective communication operations. 

\mypar{Dynamic termination} \sys{} does not yet support dynamic termination of temporal dimensions (\eg $T$ determined at runtime). Prior work~\cite{benabderrahmane2010polyhedral} has shown that this can be achieved by generating schedules that break out of infinite loops when the termination condition is met. Evaluating termination at each step, however, is expensive, because it requires a GPU/CPU transfer. We plan to explore speculative methods to reduce this overhead in future work.

\mypar{Smart transformation policies} While \sys{} proposes powerful transformations, it often lacks sophisticated policies for applying them. For example, \sys{}'s swapping augmentations are oblivious to kernel execution latencies, causing stalls when dependent operations wait for data. This can be addressed through adding constraints to the scheduling problem based on kernel profiling.
Additionally, tiling is currently triggered based on operator's point memory usage, and not on a more accurate memory usage analysis of the whole program. Finally, the tile size used in static tiling is currently set by the user, which can be suboptimal. The ideal tile size, typically the smallest size that maximizes GPU utilization~(\S\ref{sec:eval:decoding}), could instead be found through binary search.

\mypar{Optimized kernels} LLM programs today use hand-crafted, high-performance kernels, such as FlashAttention~\cite{dao2022flashattention}, which are not currently supported by \sys{}. Adding these as first-class operators to \sys{} is difficult due to their complex semantics~\cite{barham2019machine}, which prevents \sys{} from applying its transformations (vectorization, tiling, etc.). One approach is to introduce a final compiler pass, before scheduling, that identifies attention-like patterns in the SDG and lifts them~(\S\ref{sec:pdg}) into a single FlashAttention kernel operator. 


\section{Related Work}
\label{sec:rel_work}

\mypar{Memory management in DL} Prior work has explored gradient accumulation~\cite{zhao2021v} and swapping~\cite{rhu2016vdnn,huang2020swapadvisor} to reduce memory pressure. \sys transparently integrates these ideas at a program level through transformations (tiling) and scheduling (loads and evicts). Recomputation~\cite{wang2018superneurons, peng2020capuchin} of intermediate activations can also reduce peak memory usage. \sys{} does not currently offer automatic recomputation.

\mypar{Polyhedral model in DL} Tensor comprehensions~\cite{vasilache2018tensor} and Tiramisu~\cite{baghdadi2019tiramisu} optimize tensor computations through polyhedral loop fusions, tilings, and parallelization. PPCG~\cite{verdoolaege2013polyhedral} is a C-to-CUDA compiler that performs similar loop optimizations and manages memory across the GPU hierarchy. In general, these approaches use the polyhedral model for kernel-level optimizations, while \sys{} employs it for whole-program execution scheduling and memory management.

\mypar{DL compilers} TVM~\cite{chen2018tvm} and Triton~\cite{tillet2019triton} are kernel-level compilers but do not use the polyhedral model. Each introduces a domain-specific language for expressing and optimizing low-level tensor computations. MLIR~\cite{Lattner_2021} is a more general compiler infrastructure, which can be used to build domain-specific compilers, including for DL~\cite{openxla2024}. \sys{} can leverage these systems for kernel code-generation.

\mypar{Support for dynamic tensor shapes} 
JANUS~\cite{jeong2019speculative} and PyTorch~2~\cite{ansel2024pytorch} have some support for dynamic DL compilation through a combination of symbolic tensor shapes, speculative compilation, and assertions that trigger recompilation. These approaches, however, suffer from recompilation overheads when shapes change. Relax~\cite{lai2025relax} targets compilation for model inference but requires user annotations for dynamic shapes. 
Nimble~\cite{shen2021nimble} and DISC~\cite{zheng2023bladedisc} support dynamic tensor shapes through runtime adaptation. Nimble uses an interpreted approach, while DISC compiles just-in-time.
While prior approaches support dynamic shapes, \sys{} is the first system to focus on dynamic dependencies, offering a complete solution with symbolic automatic differentiation, and ahead-of-time whole-program optimization, scheduling, and memory planning with dynamic dependencies.


\section{Conclusion}

We described \sys{}, a system that combines the flexibility of eager execution with the program-wide optimizations and scheduling of graph-based execution for dynamic DL algorithms. \sys{} introduces \emph{recurrent tensors}, which add explicit \emph{temporal dimensions} to tensors that can be indexed using \emph{symbolic expressions} to express dynamic dependencies. \sys uses \emph{symbolic dependence graphs}~(SDG), to extend traditional optimizations to dynamic DL programs: it views vectorization, and tiling as transformations that move tensor dimensions from temporal to spatial and vice versa. \sys{} leverages tiling to decompose dynamic computations into static-size tiles, which enables code-generation with minimal padding. To find valid execution schedules, \sys{} uses the polyhedral model and augments SDGs to schedule memory management operations. \sys{} achieves substantial performance speed-ups, which demonstrates its potential of whole-program optimization for dynamic DL workloads.


\bibliographystyle{unsrt}
\balance
\bibliography{\jobname}

\end{document}